\documentclass[twocolumn]{aastex631}

\shorttitle{3.3\,$\mu m$ PAHs and molecular gas}
\shortauthors{Shim et al.}
\graphicspath{{./}{figures/}}

\begin{document}

\title{Possibility of using 3.3\,$\mu$m PAH luminosity as a molecular gas mass estimator}

\correspondingauthor{Hyunjin Shim}
\email{hjshim@knu.ac.kr}

\author[0000-0002-0786-7307]{Hyunjin Shim}
\affiliation{Department of Earth Science Education, Kyungpook National University, Daegu 41566, Republic of Korea} 

\author[0000-0002-3744-6714]{Junhyun Baek}
\affiliation{Korea Astronomy and Space Science Institute, Yuseong-gu, Daejeon 34055, Republic of Korea}

\author[0000-0002-6925-4821]{Dohyeong Kim}
\affiliation{Department of Earth Sciences, Pusan National University, Busan 46241, Republic of Korea}

\author[0000-0002-3560-0781]{Minjin Kim}
\affiliation{Department of Astronomy and Atmospheric Sciences, Kyungpook National University, Daegu 41566, Republic of Korea}

\author[0000-0002-4362-4070]{Hyunmi Song}
\affiliation{Department of Astronomy and Space Science, Chungnam National University, Daejeon 34134, Republic of Korea}

\author[0000-0002-5760-8186]{Gu Lim}
\affiliation{Department of Earth Sciences, Pusan National University, Busan 46241, Republic of Korea}
\affiliation{Institute for Future Earth (IFE), Pusan National University, Busan 46241, Republic of Korea}

\author{Jaejun Cho}
\affiliation{Department of Astronomy and Atmospheric Sciences, Kyungpook National University, Daegu 41566, Republic of Korea}
\author{Hayeong Jeong}
\affiliation{Department of Earth Sciences, Pusan National University, Busan 46241, Republic of Korea}
\author{Yejin Jeong}
\affiliation{Department of Earth Science Education, Kyungpook National University, Daegu 41566, Republic of Korea} 
\author{Ye-eun Kang}
\affiliation{Department of Astronomy and Atmospheric Sciences, Kyungpook National University, Daegu 41566, Republic of Korea}
\author{Dongseob Lee}
\affiliation{Department of Earth Science Education, Kyungpook National University, Daegu 41566, Republic of Korea} 
\author{Junyeong Park}
\affiliation{Department of Earth Sciences, Pusan National University, Busan 46241, Republic of Korea}
\author{Eunsuk Seo}
\affiliation{Department of Astronomy and Space Science, Chungnam National University, Daejeon 34134, Republic of Korea}
\author{Junho Song}
\affiliation{Department of Astronomy and Space Science, Chungnam National University, Daejeon 34134, Republic of Korea}
\author{Been Yeo}
\affiliation{Department of Astronomy and Atmospheric Sciences, Kyungpook National University, Daegu 41566, Republic of Korea}



\begin{abstract}
We present CO(1$-$0) observations of 50 star-forming galaxies at $0.01<z<0.35$,
for which 3.3\,$\mu$m PAH emission flux or its upper limit
is available. 
A scaling relation between 3.3\,$\mu$m PAH luminosity 
and CO(1$-$0) luminosity is established 
covering $\sim2$ orders of magnitude in total IR luminosity and CO luminosity,
with a scatter of $\sim0.23$\,dex: 
$\mathrm{log}\,L_\mathrm{3.3}/\mathrm{L}_\odot=(1.00\pm0.07)\times\mathrm{log}\,L_\mathrm{CO(1-0)}^\prime/(\mathrm{K\,km\,s^{-1}\,pc^2})+(-1.10\pm0.70)$.
The slope is near unity, allowing the use of a single value of 
$\langle\mathrm{log}\,(L_\mathrm{3.3}/L_\mathrm{CO(1-0)}^\prime)\rangle=-1.09\pm0.36~[\mathrm{L}_\odot/(\mathrm{K\,km\,s^{-1}\,pc^2})]$ 
in the conversion between 3.3\,$\mu$m PAH and CO luminosities.
The variation in the $L_\mathrm{3.3}/L_\mathrm{CO}^\prime$ ratio 
is not dependent on the galaxy properties, including total IR luminosity, 
stellar mass, and SFR excess. 
The total gas mass, estimated using dust-to-gas ratio and dust mass,
is correlated with 3.3\,$\mu$m PAH luminosity, 
in line with the prescription using $\alpha_\mathrm{CO}=0.8$-4.5 
covering both normal star-forming galaxies and starburst galaxies.
AGN-dominated galaxies tend to have a lower $L_\mathrm{3.3}/L_\mathrm{CO}^\prime$
than non-AGN galaxies, which needs to be investigated further with an increased sample size.
The established $L_\mathrm{3.3}$-$L_\mathrm{CO}^\prime$ correlation 
is expected to be applicable to wide-field near-infrared spectrophotometric surveys
that allow the detection of 3.3\,$\mu$m emission from numerous low-redshift galaxies. 
\end{abstract}




\section{Introduction} \label{sec:intro}

Star formation is one of the most important processes in  
galaxy formation, evolution, and growth. 
An overall history of cosmic star formation, which has been continuously declining
since $z\sim2$ \citep{2014ARA&A..52..415M}, is consistent 
with the history of global stellar mass density 
while the stellar mass is described
as an integration of the past star formation rate (SFR). 
Observed scaling relations between stellar mass, SFR, and gas-phase metallicity 
\citep[e.g.,][]{2004ApJ...613..898T, 2010MNRAS.408.2115M, 2011ApJ...742...96W, 2014ApJS..214...15S}
suggest relatively tight correlations between these parameters,
implying that the star formation is regulated by molecular gas 
which is a fuel to star formation, 
while a universal equilibrium exists between physical processes
including gas supply, star formation, and gas expulsions as well as recycling
\citep{2010ApJ...718.1001B, 2013ApJ...772..119L}.

Based on the well-known tight correlations between SFR and gas surface density 
for star-forming galaxies \citep{1959ApJ...129..243S, 1998ApJ...498..541K}, 
the decline of global SFR density 
could result from the decrease of gas content in galaxies \citep{2013ApJ...762L..31B},
while the gas mass fraction in high-redshift star-forming galaxies 
is observed to be higher than that in local spiral galaxies 
\citep{2010ApJ...713..686D, 2010Natur.463..781T, 2011MNRAS.415...32S}.
The amount of young stars formed per unit molecular gas mass 
($\equiv\mathrm{SFR}/M_\mathrm{gas}$), defined as a star formation efficiency, 
is known to be elevated in starburst galaxies
\citep{1997ApJ...478..144S, 2010MNRAS.407.2091G, 2012A&A...539A...8G}. 
While star formation in galaxies could be classified into 
two different modes, one for normal main-sequence mode and 
the other for starburst mode \citep{2011A&A...533A.119E}, 
the star formation efficiency may vary 
locally within galaxies, due to the varying giant molecular cloud formation 
in different interstellar medium environments \citep{2008AJ....136.2782L}. 
To get a full understanding of the characteristics of 
star-forming galaxy populations at different epochs 
and to trace their evolution history, 
it is necessary to study the star formation efficiency
through the measurement of SFR as well as the estimates of molecular gas mass. 

Due to the non-polar nature of the hydrogen molecule ($\mathrm{H_2}$), 
molecular gas in galaxies is in general traced using indirect tracers, 
one of the most commonly employed being the rotational transition line 
of the carbon monoxide \citep[CO; e.g.,][]{1991ARA&A..29..581Y, 1995ApJS...98..219Y, 2011MNRAS.415...32S, 
2013MNRAS.429.3047B}.
The observed CO luminosity is converted to $\mathrm{H_2}$ mass 
using the CO-to-$\mathrm{H_2}$ conversion factor ($\alpha_\mathrm{CO}$),
which is known to vary for galaxies with different physical parameters 
such as SFR, metallicity, and gas dynamics
\citep{2011ApJ...737...12L, 2013ARA&A..51..207B}.
To reduce uncertainties in molecular gas estimates 
due to $\alpha_\mathrm{CO}$ uncertainties 
and to complement gas estimates that require CO line detection, 
other strategies have been suggested including 
the use of far-infrared (FIR) fluxes in estimating dust mass and scaling it to gas mass
by applying dust-to-gas ratio \citep{2014ApJ...783...84S, 2016ApJ...820...83S, 2016A&A...587A..73B}, 
which also has inherent uncertainties 
due to dust modeling uncertainty and the metallicity dependence of 
the dust-to-gas ratio \citep{2014A&A...563A..31R}. 

Recently, \citet{2019MNRAS.482.1618C} suggested that
there exists a universal scaling relation between 
mid-infrared (6.2 and 7.7\,$\mu$m) polycyclic aromatic hydrocarbon (PAH) luminosity 
and CO luminosity for star-forming galaxies. 
PAH molecules in the photo-dissociation regions,
excited by the far-ultraviolet stellar radiation, 
produce strong emission features at 3.3, 6.2, 7.7, 8.6, 11.3, and 12.7\,$\mu$m.
Therefore, PAH emission has been suggested to be 
a qualitative and quantitative tracer of star formation in a wide range of galaxies
\citep{2004ApJ...613..986P, 2016ApJ...818...60S}. 
Considering the tight correlation between SFR and molecular gas mass, 
it is naturally expected that PAH emission and molecular gas content would show a correlation.
The correlation between PAH and CO emission 
is being investigated even in spatially resolved scales for nearby galaxies
\citep{2023ApJ...944L..11C, 2023ApJ...944L..10L}.

While the shortest wavelength PAH emission feature, 
i.e., 3.3\,$\mu$m PAH emission, 
is much weaker in terms of flux density 
thus less investigated so far compared to longer PAH emission features, 
the potential of using 3.3\,$\mu$m PAH 
in the study of star formation in nearby galaxies 
\citep[through all-sky near-infrared spectrophotometric surveys such as 
SPHEREx;][]{2018arXiv180505489D}
is promising due to the large sample size. 
Since the 3.3\,$\mu$m PAH is the shortest among PAH features, 
it serves as a tracer of dust-free star formation in star-forming galaxies at 
the reionization era (using mid-infrared spectroscopic instruments such as JWST MIRI).    
To use 3.3\,$\mu$m PAH emission either in SFR estimation 
or in molecular gas mass estimation, 
an assessment of whether there exists a correlation between 3.3\,$\mu$m PAH 
and CO luminosities should be preceded 
using samples with both PAH and CO observations. 

In this work, we establish the scaling relation 
between the 3.3\,$\mu$m PAH luminosity and gas content (CO luminosity and molecular gas mass)
using new single-dish CO(1$-$0) observations of 50 objects including both star-forming galaxies and AGN.
The sources, i.e., targets of new CO observations, are selected from literature
listing 3.3\,$\mu$m PAH flux measurements (Section~\ref{sec:sample}). 
By adding measurements from 
our new observations (Sections~\ref{sec:observation} and \ref{sec:datareduction}) 
to the existing data (Section~\ref{sec:compilation}), 
the number of objects for which both 3.3\,$\mu$m and CO luminosities are available
has increased by a factor of $\sim1.6$. 
We present the constructed $L_\mathrm{3.3}$-$L_\mathrm{CO}^\prime$ correlation 
in Section~\ref{sec:corr} and 
discuss the molecular gas mass estimation from 3.3\,$\mu$m PAH in Section~\ref{sec:gasmass}. 
In addition to the overall $L_\mathrm{3.3}$-$L_\mathrm{CO}^\prime$ correlation, 
comparison between AGN-dominated galaxies 
and non-AGN galaxies are investigated in Section~\ref{sec:agn}.
Throughout the paper, we use 
the flat $\Lambda$CDM cosmology model
with $\Omega_\mathrm{m, 0}=0.3$, $H_0=70$\,km\,s$^{-1}$\,Mpc$^{-1}$.
When using values from different literature,
the differences in cosmological parameters are taken into account.

\section{Data}   \label{sec:data}

\subsection{3.3\,$\mu$m-PAH-selected galaxies}    \label{sec:sample}

To explore the relationship between 3.3\,$\mu$m PAH luminosity and CO luminosity, 
we constructed a 3.3\,$\mu$m-PAH-selected sample of 180 sources at $0.01<z<0.35$
by compiling the previous 2.5-5\,$\mu$m spectroscopic observations
from the infrared satellite 
AKARI \citep{2013PASJ...65..103Y, 2019PASJ...71...25K, 2020ApJ...905...55L}. 
The 2.5-5\,$\mu$m spectra were obtained with
the IRC \citep[Infrared Camera;][]{2007PASJ...59S.401O}, 
utilizing the near-infrared grism (NG) mode for point source spectroscopy
that yields the spectroscopic resolution of $R\sim120$ \citep{2007PASJ...59S.411O}.
The observations were made with a $1\arcmin\times1\arcmin$ aperture, 
thus the 3.3\,$\mu$m PAH luminosity measured from the AKARI/IRC spectrum
corresponds to the value integrated over the entire galaxy if the galaxy is smaller than $1\arcmin$.
The target selection varied across different IRC programs,
including mid-infrared excess sources with higher 9\,$\mu$m flux density compared 
to the $K_s$-band flux density \citep{2011A&A...529A.122O},
local luminous infrared galaxies (LIRGs; $\mathrm{log}\,(L_\mathrm{IR}/\mathrm{L}_\odot)>11$) 
and ultraluminous infrared galaxies (ULIRGs; $\mathrm{log}\,(L_\mathrm{IR}/\mathrm{L}_\odot)>12$)
\citep{2008PASJ...60S.489I, 2010ApJ...721.1233I}, 
and unusual galaxies such as low-metallicity dwarf galaxies \citep{2020ApJ...905...55L}.
Therefore, the 3.3\,$\mu$m-PAH-selected sample consists of 
heterogeneous galaxy populations, 
covering a wide range of total infrared luminosities ($10^{9-12.5}\,\mbox{L}_\odot$),
including both AGN-dominated and star formation-dominated systems. 

For the sources from \citet{2020ApJ...905...55L},
the luminosities of PAH emission features at other wavelengths 
(e.g., 6.2, 7.7, and 11.3\,$\mu$m) are also available 
since the sources are from the constructed list of `bright-PAH' galaxies, 
which have spectra with full coverage from 2.5 to 38\,$\mu$m 
by combining the Spitzer \citep[IDEOS;][]{2016MNRAS.455.1796H, 2022ApJS..259...37S} and AKARI spectra. 
To include the AKARI sources that are not matched to the IDEOS catalog, 
we used the works by \citet{2013PASJ...65..103Y} and \citet{2019PASJ...71...25K}. 
Note that the data reduction package versions are different for different works, e.g.,
``IRC Spectroscopy Toolkit for Phase 3 data Version 20090211'' \citep{2013PASJ...65..103Y}
and ``IRC Spectroscopy Toolkit for Phase 3 data Version 20170225'' \citep{2020ApJ...905...55L}
for Phase 3 data, 
``IRC Spectroscopy Toolkit Version 20090211'' \citep{2013PASJ...65..103Y}
and ``IRC Spectroscopy Toolkit Version 20080528'' \citep{2020ApJ...905...55L}
for Phases 1 and 2 data. The major updates on the data reduction package were related to
the revision of the wavelength and spectral response calibrations of
grism\footnote{https://www.ir.isas.jaxa.jp/AKARI/Observation/support/IRC/},
thus the 3.3\,$\mu$m PAH luminosity values of a single source presented in
multiple works differ from each other due to flux calibration uncertainties
and different aperture sizes used in spectra extraction.
In addition to these, different works used different treatments for dust attenuation correction.
\citet{2020ApJ...905...55L} present 
attenuation-corrected 3.3\,$\mu$m PAH luminosity based on the two different dust geometry assumptions,
mixed and obscured continuum. 
The 3.3\,$\mu$m PAH luminosity values presented by 
\citet{2013PASJ...65..103Y} and \citet{2019PASJ...71...25K} are not corrected for dust attenuation.

The numbers of sources compiled from \citet{2020ApJ...905...55L}, \citet{2013PASJ...65..103Y},
and \citet{2019PASJ...71...25K} are 112, 95, and 14, respectively.
For 41 sources from \citet{2020ApJ...905...55L}
that overlap with sources in \citet{2013PASJ...65..103Y},
the average difference in $\mathrm{log}\,L_{3.3}$ values is
$0.055\pm0.145$ and $0.237\pm0.198$\,dex
when obscured continuum and mixed geometry assumptions were used, respectively. 
In the case of the 3.3\,$\mu$m PAH luminosity values corrected for mixed dust geometry, 
$\Delta \mathrm{log}\,L_{3.3}$ (discrepancy from the value without attenuation correction) 
tends to correlate with the 3.3\,$\mu$m PAH luminosity.
Therefore, to reduce the uncertainties 
generated by the systematic difference between different $\mathrm{log}\,L_{3.3}$ values,
we used 3.3\,$\mu$m PAH luminosity 
corrected for dust attenuation based on the obscured continuum geometry,
rather than the mixed geometry, for 112 sources from \citet{2020ApJ...905...55L}. 
Among 54 additional sources from \citet{2013PASJ...65..103Y},
18 sources are not detected in 3.3\,$\mu$m, 
as well as 12 out of 14 sources in \citet{2019PASJ...71...25K}.
For these, 3.3\,$\mu$m upper limits were compiled.

The multi-wavelength photometry data points covering UV, optical, NIR, MIR, and FIR regimes
were compiled for 3.3\,$\mu$m-PAH-selected sources 
by cross-identifying the positions of sources 
in the archival GALEX \citep{2017ApJS..230...24B}, SDSS \citep{2015ApJS..219...12A}, 
2MASS \citep{2006AJ....131.1163S}, UKIDSS \citep{2007MNRAS.379.1599L},
WISE \citep{2010AJ....140.1868W}, 
IRAS \citep{1992ifss.book.....M}, 
AKARI \citep{2010yCat.2298....0Y}, 
and Herschel/SPIRE \citep{2024yCat.8112....0H} source catalogs.
The search radius was $3\arcsec$ in optical and $6\arcsec$ in other wavelengths,
and the total (integrated) flux density was adopted when the source was spatially resolved.

Using the photometry compilation, the physical properties of each source 
including stellar mass, SFR, and dust mass
were derived through spectral energy distribution (SED) fitting. 
The SED fitting was performed using \textsc{cigale}\footnote{https://cigale.lam.fr/2020/06/29/version-2020-0/}
\citep[Code Investigating GALaxy Emission,][]{2009AandA...507.1793N, 2019A&A...622A.103B},
with similar configurations as in \citet{2023AJ....165...31S}. 
The \texttt{bc03} \citep{2003MNRAS.344.1000B} stellar population model 
with the delayed star formation history (\texttt{sfhdelayed}) 
was used to estimate stellar mass and SFR.
The mid-infrared to far-infrared SED was fitted using the \texttt{dl2014} \citep{2014ApJ...780..172D} 
dust emission model with varying PAH mass fraction, stellar radiation field intensity, 
and the fraction of dust mass exposed to starlight. 
While the AGN component contribution to the SED is also estimated 
using \texttt{fritz2006} model \citep{2006MNRAS.366..767F}, 
we did not use the AGN fraction in the type classification 
among 3.3\,$\mu$m-PAH-selected sources
since the AGN classification based on the SED fitting requires
fine sampling of the mid-infrared SED \citep[e.g.,][]{2020MNRAS.499.4068W}.
Instead, we used either optical spectra (from SDSS) or MIR colors (from WISE) to 
classify sources into AGN-dominated and star formation-dominated subgroups, 
based on the existence of broad lines 
and/or ($W1-W2$) vs. ($W2-W3$) colors (using the criteria of
\citealt{2012MNRAS.426.3271M}) that reflect the warm dusty torus
\citep{2011ApJ...735..112J, 2012MNRAS.426.3271M}.

\begin{figure}
\epsscale{1.15}
\plotone{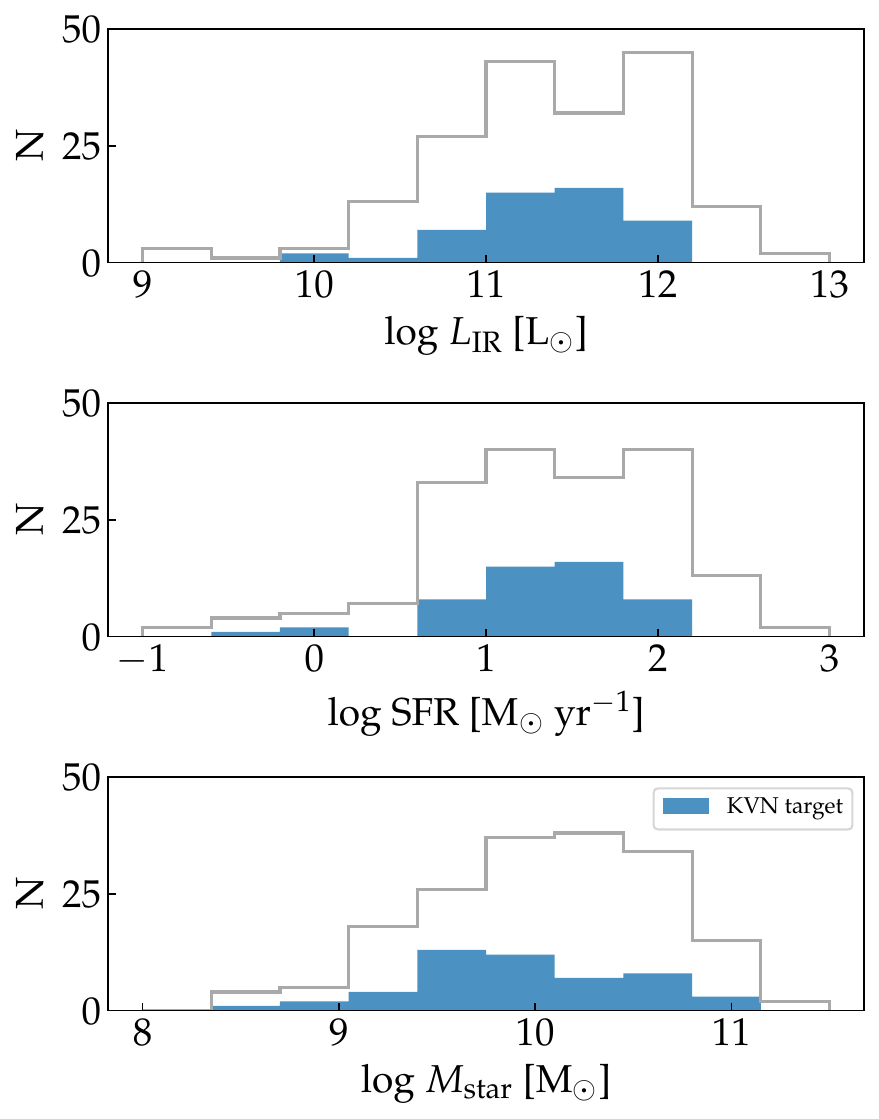}
\caption{Total infrared luminosity (top), star formation rate (middle),
and stellar mass (bottom) distributions of the 3.3\,$\mu$m-selected sources.
Shaded histograms correspond to 50 targets in our KVN CO observation programs.}
\label{fig:sample}
\end{figure}

\begin{figure}
\epsscale{1.15}
\plotone{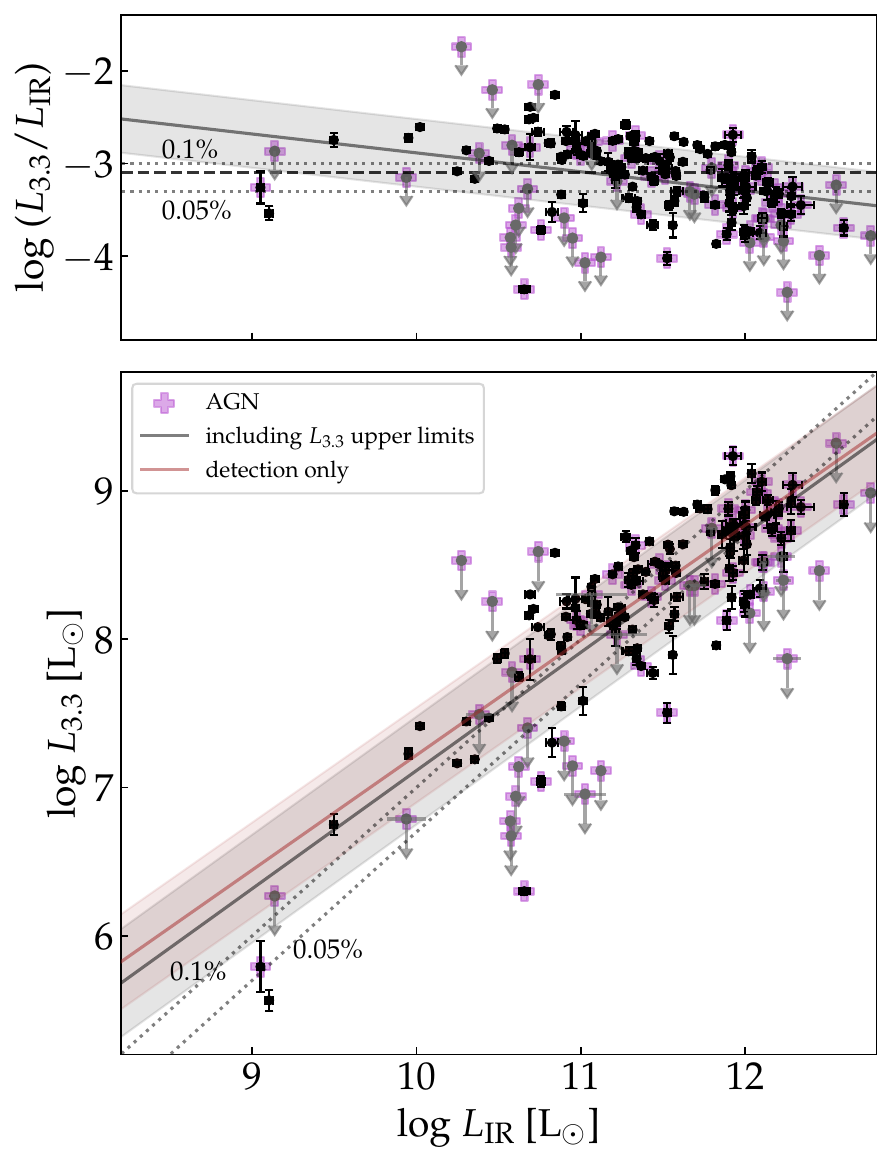}
\caption{Comparison between 3.3\,$\mu$m PAH luminosity and total IR luminosity.
Potential AGN, identified by either broad hydrogen recombination lines or MIR colors, 
are marked with crosses. 
Solid lines depict the best-fit linear relationship between 
the $\mathrm{log}\,L_\mathrm{3.3}$ vs. $\mathrm{log}\,L_\mathrm{IR}$,
while shaded regions represent
the intrinsic scatter of the fits.
The dashed line in the upper panel is a horizontal line at $-3.1$, 
corresponding to
the mean $\mathrm{log}\,(L_{3.3}/L_\mathrm{IR})$ value for the entire sample
(150 sources, excluding 30 sources with 3.3\,$\mu$m upper limits).
Dotted lines represent $L_{3.3}/L_\mathrm{IR}=$\,0.1\,\% and 0.05\,\%, 
a typical range of 3.3\,$\mu$m PAH luminosity \citep{2013PASJ...65..103Y, 2020ApJ...905...55L}.
}
\label{fig:sample33lir}
\end{figure}

Figure~\ref{fig:sample} shows the total IR luminosity (integrated over 8-1000\,$\mu$m), 
SFR, and stellar mass distributions of the 
3.3\,$\mu$m-PAH-selected sources. 
Most (over 70\,\%) sources are classified as either LIRGs or ULIRGs,
corresponding to an SFR of 10-1000\,$\mathrm{M}_\odot\,\mathrm{yr}^{-1}$. 
In the SFR-stellar mass diagram, 
these galaxies are located on average $\sim1$\,dex 
above the main sequence of star-forming galaxies at $z<0.3$
\citep{2014ApJS..214...15S, 2023MNRAS.519.1526P}.

Previous works have discussed that prominent MIR (e.g., 6.2\,$\mu$m and 7.7\,$\mu$m) 
PAH features get weakened as the IR luminosity increases,
with the typical L(PAH)-L(IR) relation breaking  
at around $\mathrm{log}\,(L_\mathrm{IR}/\mathrm{L}_\odot)\sim11.5$ 
for local galaxies \citep{2019MNRAS.482.1618C}
and $\mathrm{log}\,(L_\mathrm{IR}/\mathrm{L}_\odot)\sim12.5$
for higher-redshift galaxies \citep{2008ApJ...675.1171P, 2013ApJ...772...92P, 2016ApJ...818...60S}. 
In the case of 3.3\,$\mu$m PAH, \citet{2020ApJ...905...55L} showed that
the luminosity ratio $L_{3.3}/L_\mathrm{IR}$ is nearly constant at
below $\mathrm{log}\,(L_\mathrm{IR}/\mathrm{L}_\odot)=11.2$.  
Figure~\ref{fig:sample33lir} shows the correlation between total IR luminosity 
and 3.3\,$\mu$m PAH luminosity. The best-fit linear correlation 
between $\mathrm{log}\,L_{3.3}$ and $\mathrm{log}\,L_\mathrm{IR}$ 
(black solid line in Figure~\ref{fig:sample33lir}, Table~\ref{tab:fit})
is derived using the approach described in \citet{2007ApJ...665.1489K}, 
and is expressed as:

\begin{equation}
   \mathrm{log}\,L_{3.3} = (0.80\pm0.04)\times\mathrm{log}\,L_\mathrm{IR}+(-0.85\pm0.51).
   \label{eq:LIR33}
\end{equation}

By excluding points with $L_{3.3}$ upper limits, i.e., censoring these points, 
the derived relationship (brown solid line in Figure~\ref{fig:sample33lir}) is comparable to the original one 
within an intrinsic scatter ($\sigma\sim0.36$\,dex).
The sub-unity (0.80) slope between $\mathrm{log}\,L_{3.3}$ and $\mathrm{log}\,L_\mathrm{IR}$ 
is comparable to that of 
the cases of $\mathrm{log}\,L_{6.2}$ and $\mathrm{log}\,L_{7.7}$ 
\citep{2019MNRAS.482.1618C}. 
The derived trend remains almost the same if AGN are excluded (Table~\ref{tab:fit});
note that it cannot be ruled out that such no change is due to the small number of AGN compared to non-AGN. 
The main reason causing sub-unity slope is the decrease of $\mathrm{log}\,(L_{3.3}/L_\mathrm{IR})$
at $\mathrm{log}\,(L_\mathrm{IR}/\mathrm{L}_\odot) >11.2$;
the median of the $(L_{3.3}/L_\mathrm{IR})$ decreases as the IR luminosity increases, 
consistent with the findings of
\citet{2020ApJ...905...55L} (using $L_{3.3}$ corrected for dust attenuation
assuming obscured continuum geometry).

\subsection{New CO(1$-$0) observations}  \label{sec:observation}

The KVN (Korean VLBI Network) is a long baseline interferometry network in Korea 
that consists of three 21\,m radio antennas 
located at Yonsei (KYS), Ulsan (KUS), and Tamna (KTN) sites \citep{2011PASP..123.1398L},
with a fourth site added in mid-2024 at Pyeongchang (KPC).
Besides its capability for performing interferometry observations,
the KVN also offers the single-dish (SD) observing mode.

We carried out CO(1$-$0) observations 
of 50 3.3\,$\mu$m-PAH-selected galaxies (Table~\ref{tab:summary})
using the three antennas with the SD observing mode,
from 2023 December to 2024 March.
The targets for the KVN observations were selected based on their visibility. 
The beam size (half power beam width) of the KVN 21\,m antennas
at $\sim115$\,GHz is $\sim30\arcsec$, so we prioritized sources with 
optical radii ($D_\mathrm{25}$) 
smaller than $60\arcsec$, twice the beam size. 
However, due to the proximity of the sources where 3.3\,$\mu$m PAHs are available, 
$\sim30$\,\% of the KVN sources (16 out of 50)
have optical radii larger than $60\arcsec$,
raising the possibility that their CO measurements could be a lower limit. 
The IR luminosity and SFR distribution of the KVN targets are similar 
to those of the parent 3.3\,$\mu$m sample (Figure~\ref{fig:sample}),
while the stellar masses of the KVN targets are slightly lower 
as the smaller-sized source was prioritized. 

All sources were observed using the W-band (80-116\,GHz)
with the central frequency of the GPU spectrometer
tuned to the expected CO(1$-$0) frequency of each target
and the backend (wide-field sampler, OCTAD) bandwidth of 1024\,MHz. 
The observations were executed in a position-switching mode, 
i.e., consecutively having 10 seconds of off-source exposure 
after each 10 seconds of on-source exposure
for proper removal of the sky background. 
The actual on-source integration time varies
for different objects (between 0.5 and 4.5\,hours), 
since we stopped the integration if the $S/N>5$ was reached
for the CO emission line 
with a spectral resolution of 10\,km\,s$^{-1}$.
During the observations, pointing calibrations were performed
every $\sim3$\,hours by observing position reference sources.
Focus adjustments were made every $\sim6$\,hours.  

\begin{figure*}
\epsscale{1.05}
\plotone{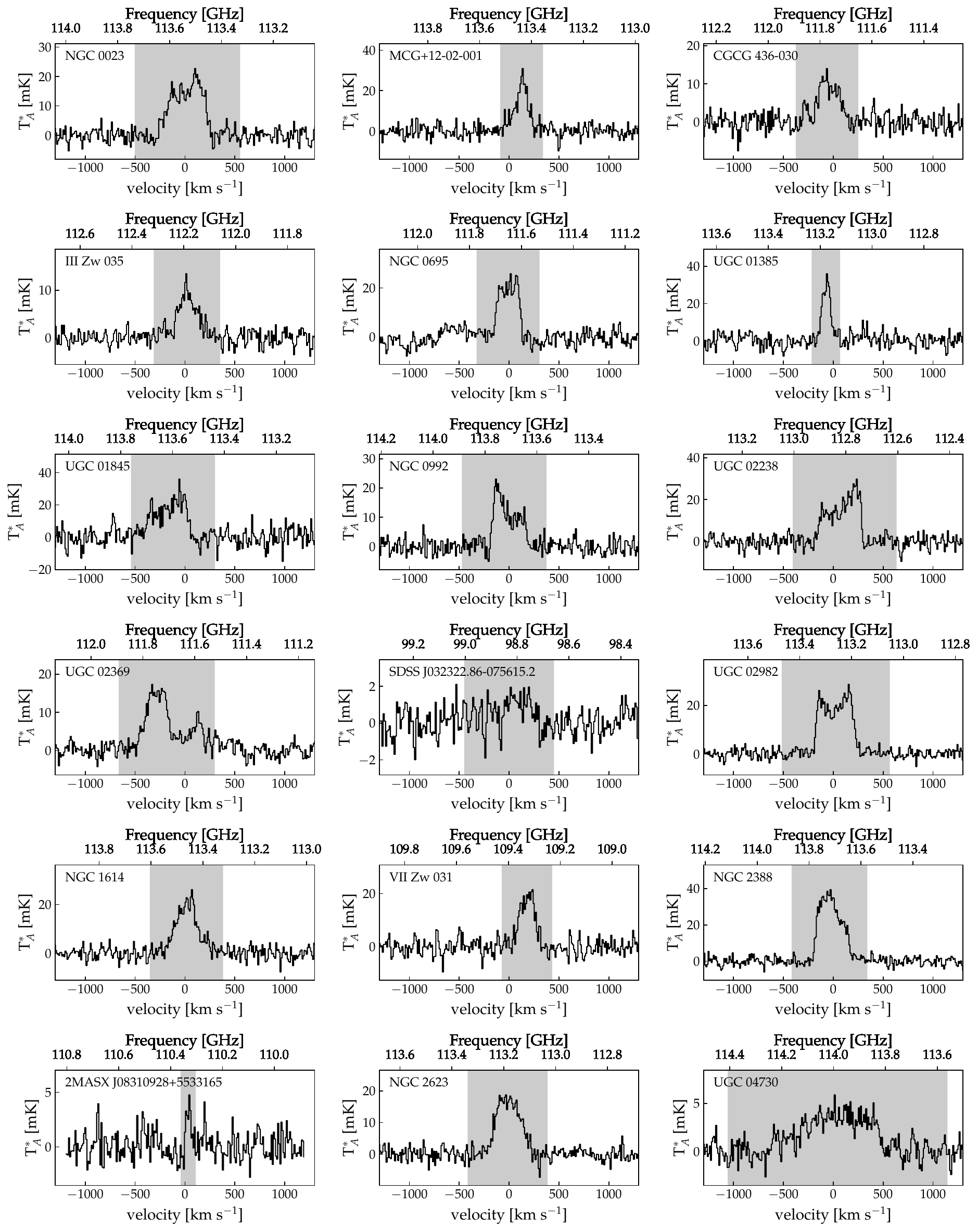}
\caption{\label{fig:spec}
 CO(1$-$0) spectra of 50 3.3\,$\mu$m-PAH-selected objects 
 obtained from the KVN SD observations. 
 All spectra are smoothed to a velocity resolution of $\sim10$\,km\,s$^{-1}$.
The gray shaded regions indicate the range of velocities used to compute the 
velocity-integrated flux density (equivalent to three times line width).
}
\end{figure*}

\renewcommand{\thefigure}{\arabic{figure} (Continued)}
\addtocounter{figure}{-1}

\begin{figure*}
\epsscale{1.05}
\plotone{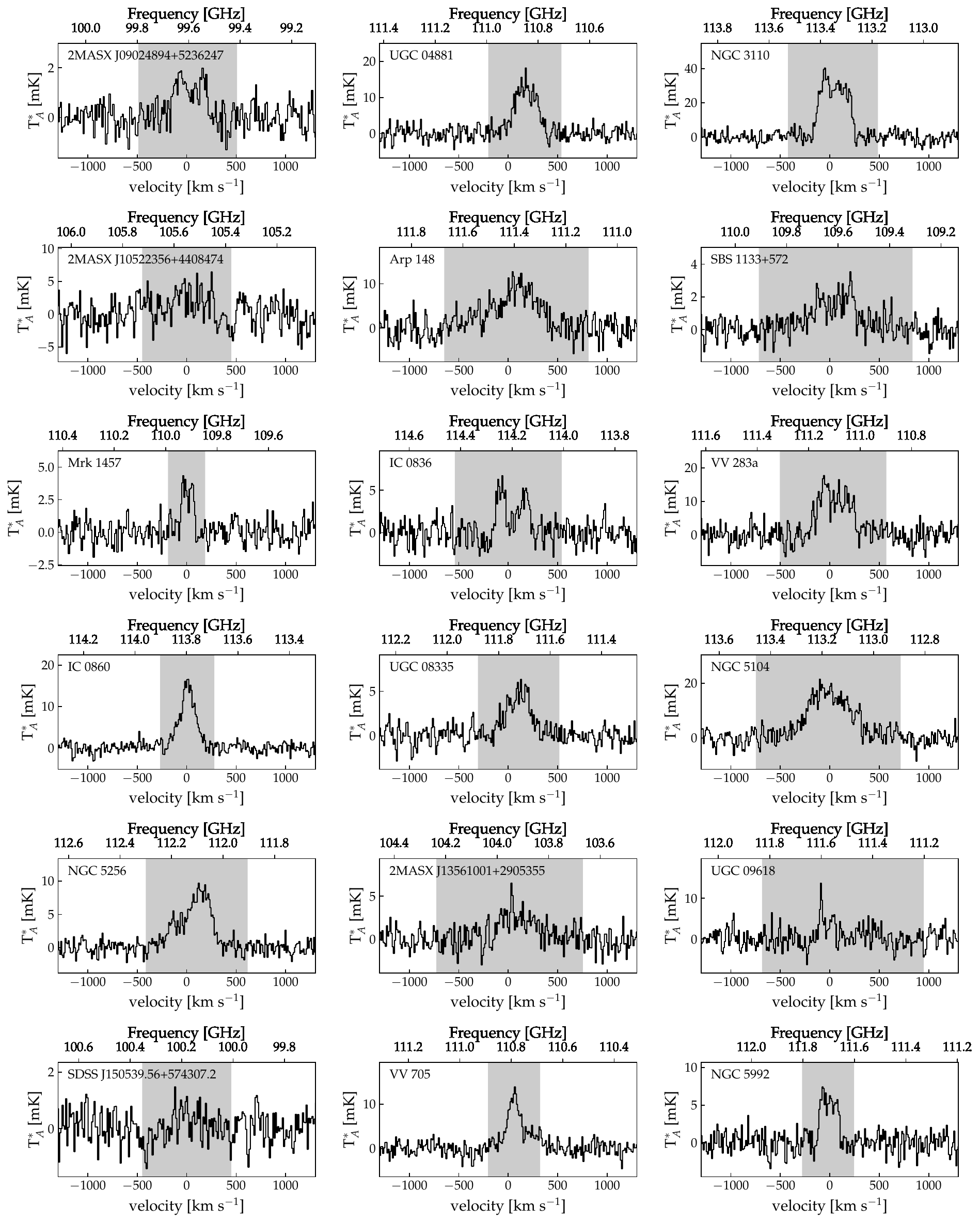}
\caption{}
\end{figure*}

\renewcommand{\thefigure}{\arabic{figure} (Continued)}
\addtocounter{figure}{-1}

\begin{figure*}
\epsscale{1.05}
\plotone{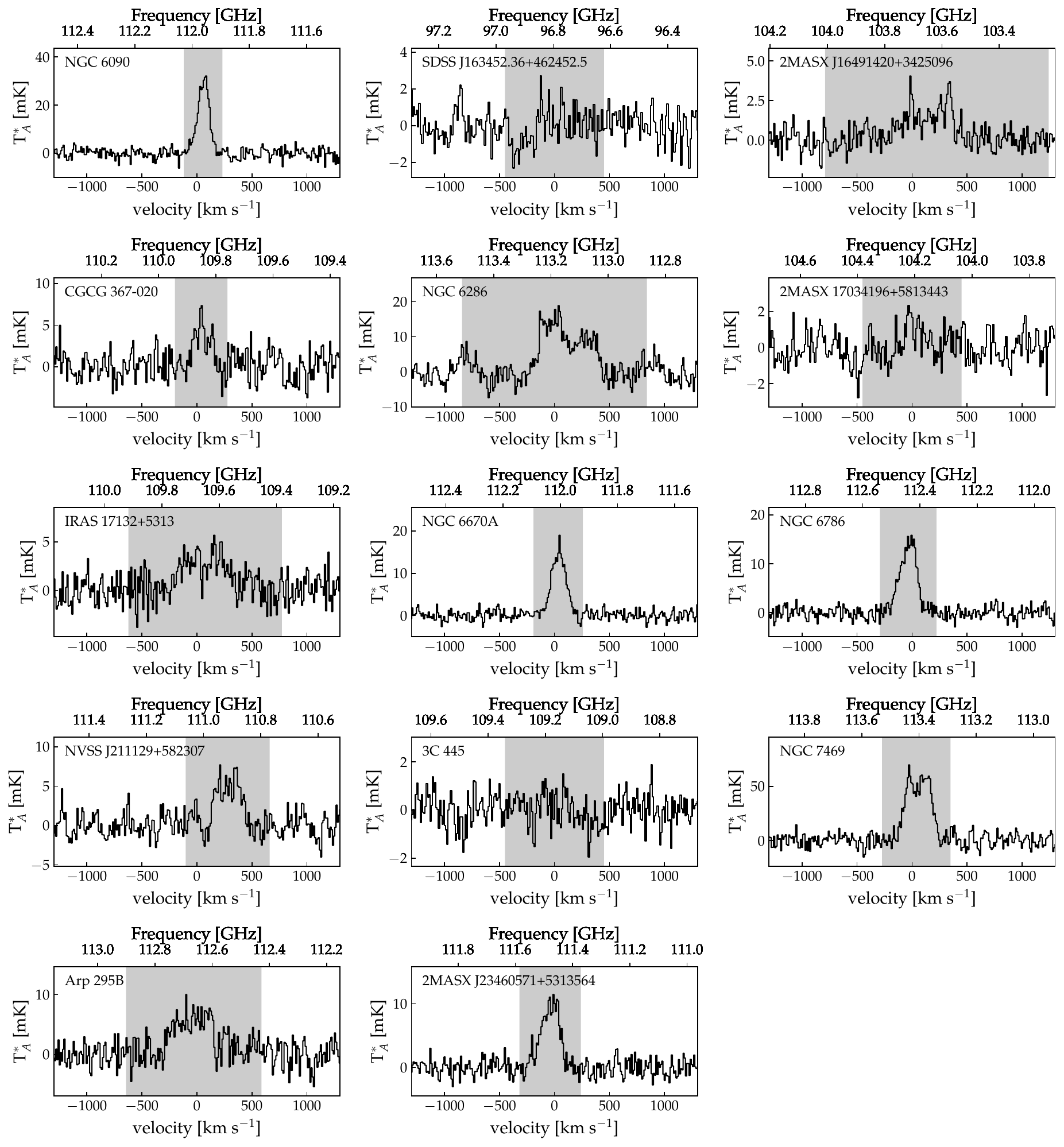}
\caption{}
\end{figure*}

\renewcommand{\thefigure}{\arabic{figure}}

\begin{figure}
\epsscale{1.15}
\plotone{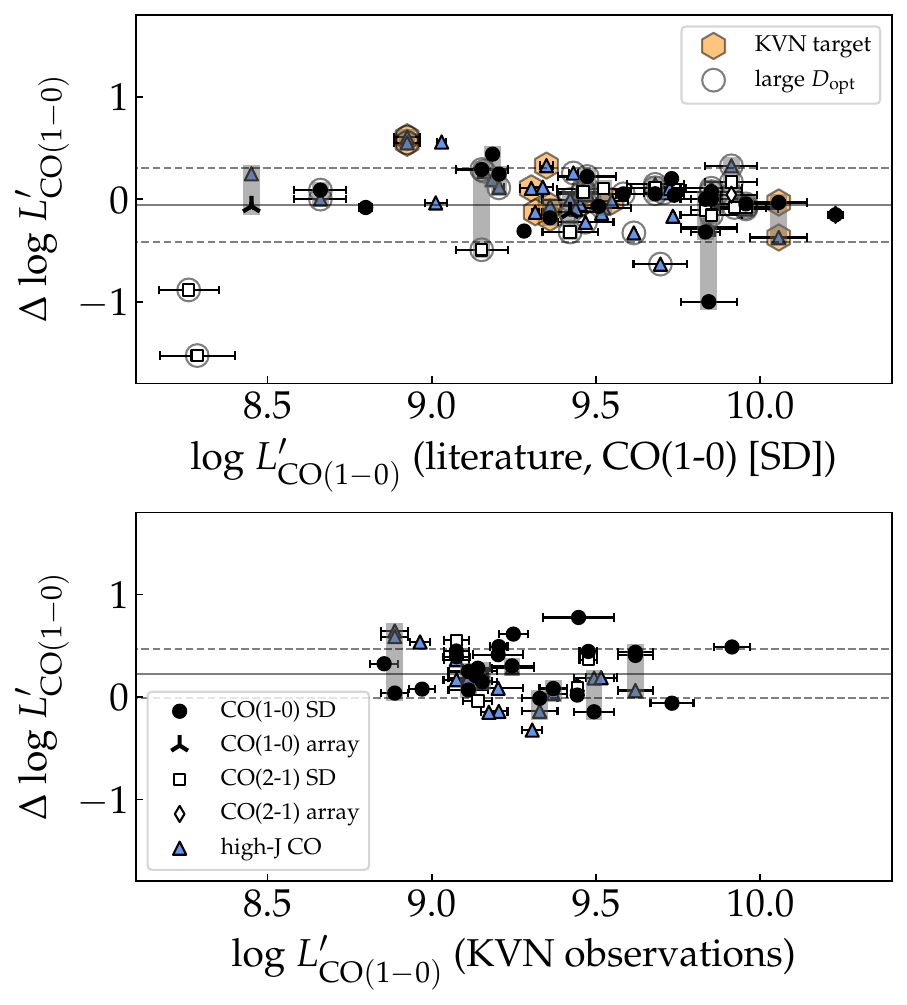}
\caption{\label{fig:fluxcomp}
(Top) Comparison between the CO luminosities from different literature.
Different CO transitions and observation types are 
plotted as different symbols (see legend in bottom panel). 
KVN targets and large sources are marked. 
Vertically connected points (with thick gray lines) indicate one identical source.
Horizontal solid and dashed lines indicate mean difference
and standard deviation ($-0.054\pm0.362$). 
(Bottom) Difference of $\mathrm{log}\,L_\mathrm{CO(1-0)}^\prime$ 
between our KVN observations and other literature. 
Again, the mean difference and standard deviation ($0.226\pm0.239$)
are indicated as horizontal lines.
}
\end{figure}

\subsection{Data reduction}    \label{sec:datareduction}

Spectral data reduction was done using 
the IRAM GILDAS/CLASS\footnote{http://www.iram.fr/IRAMFR/GILDAS/} 
software \citep{2013ascl.soft05010G}.
We inspected individual spectra for a given observing target 
to exclude low-quality scans, such as those with highly fluctuating baselines
due to bad weather conditions.
Then the remaining scans were averaged using sigma weighting, i.e., 
weighted by the inverse square of the rms noise of each individual spectrum. 
The rms noise was estimated in each individual spectrum 
by assuming a linear baseline 
outside a window of [$-600$, $+600$]\,km\,s$^{-1}$.  
The averaged spectra were smoothed to the velocity resolution of $\sim10$\,km\,s$^{-1}$. 

To convert antenna temperature ($T_A^*$) in K into flux density Jy 
in the averaged spectra, 
we used the Degree Per Flux density Unit (DPFU) values described in the 
2023 KVN status report\footnote{https://radio.kasi.re.kr/status\_report/}
(0.0572, 0.0685, 0.0548 K Jy$^{-1}$ for KYS, KUS, KTN site, respectively).
The CO(1$-$0) line luminosities, $L_\mathrm{CO}^\prime$, were estimated in units of 
[K\,km\,s$^{-1}$\,pc$^{2}$] using the following equation 
\citep{1997ApJ...478..144S, 2005ARAandA..43..677S}:

\begin{equation}
L_\mathrm{CO}^\prime = 3.25\times10^7 S_\mathrm{CO}\Delta v ~\nu_\mathrm{obs}^{-2} \,D_\mathrm{L}^2 (1+z)^{-3}, 
\end{equation}

\noindent where $S_\mathrm{CO}\Delta v$ is the velocity-integrated flux density
in Jy km s$^{-1}$, 
$\nu_\mathrm{obs}$ is the observed frequency, 
and $D_\mathrm{L}$ is the luminosity distance in Mpc. 

Figure \ref{fig:spec} shows the CO(1$-$0) spectra of the 50 targets
in our KVN observation programs.  
The line is considered to be detected if the peak antenna temperature is above $4\sigma$.
We detected a CO(1$-$0) line for 44 (88\,\%) out of 50 sources.
Some sources clearly show a double-peaked line profile, 
which is often asymmetric, 
reflecting the rotational motion of the molecular gas. 
For sources where the CO line is approximated as a single Gaussian, 
we consider the FWHM of the Gaussian profile as the line width ($\Delta\mathrm{W}$).
In the case of a double-peaked line profile, 
we assume the width of the double-horned shape to be the line width.
The velocity-integrated CO flux density $S_\mathrm{CO}\Delta v$
was derived by integrating the area under the line 
in the integration range of [$-1.5\,\Delta\mathrm{W}$, $+1.5\,\Delta\mathrm{W}$]
from the line center. By changing the integration range to [$-\Delta\mathrm{W}$, $+\Delta\mathrm{W}$],
the flux density decreases by 1 to 9\,\%,
which is minimal compared to the scatter in CO measurements from different transitions
and observing methods (Section~\ref{sec:compilation}, Figure~\ref{fig:fluxcomp}).

The $S_\mathrm{CO}\Delta v$, line width (FWHM; $\Delta\mathrm{W}$), 
and other observation-related properties such as the coordinate, redshift derived from the CO emission line, 
$\mathrm{log}\,L_{3.3}$, $\mathrm{log}\,L_\mathrm{CO(1-0)}^\prime$, and the rms noise 
are listed in Table~\ref{tab:summary}. 
Note that the coordinates given in Table~\ref{tab:summary}
are coordinates of the KVN pointings, which can be slightly offset from that of 3.3\,$\mu$m PAH pointings
especially in sources that is a member of an interacting/merging system (see table notes). 
For six sources without CO detections, the $3\sigma$ flux upper limits are estimated
assuming a line width of 300\,km\,s$^{-1}$.

\subsection{CO from literature}     \label{sec:compilation}

We compiled the existing CO emission line measurements 
(either from single dish or interferometry observations) 
for 3.3\,$\mu$m-PAH-selected sources from numerous literature 
\citep[Table \ref{tab:literature},][]{1995ApJS...98..219Y, 2015AandA...578A..95I, 2015PASJ...67...36M, 
2016ApJ...833L...6C, 2017ApJS..230....1L, 2017ApJ...844...96Y, 
2019MNRAS.482.1618C, 2019AandA...628A..71H, 2020ApJS..247...15S, 2021ApJS..257...64K, 
2022AandA...664A..60B, 2022AandA...668A..45L, 2023AandA...673A..13M}.
Many sources were targeted by more than one observing program, 
thus we have at most five CO measurements from the literature
for a single source. 
In the case of the interferometric CO observations
that provide spatially resolved CO emission, 
we utilized the sum of the flux densities from individual components 
to account for the integrated value from the entire galaxy.  
For higher CO transitions (e.g., $J>1$ with $J-(J-1)$ transition),
we converted the observed values to those for (1$-$0) transition 
using the values measured for submillimeter galaxies
\cite[e.g., $r_{21/10}=0.84\pm0.13$;][]{2013MNRAS.429.3047B}
considering the large SFR and IR luminosity of these galaxies. 
Note, however, since the ratios between different CO transition lines 
heavily depend on the physical conditions of the interstellar gas 
and the processes related to the star formation, 
the scatter in the line ratios is highly significant 
(e.g., $r_{21/10}$ ranging from 0.33 to 1.47) for LIRGs. 
Therefore, caution is needed when including high-$J$ CO transitions 
in the construction of the PAH-CO calibration formula.

In the following analysis, if multiple CO measurements (from multiple literature) are available 
for one source, we adopt the values in the following order of preference: 
CO(1$-$0) SD (63 sources), CO(1$-$0) array (9 sources), CO(2$-$1) SD (4 sources), 
CO(2$-$1) array (7 sources), and high-$J$ CO transitions (18 sources). 
Among these, 26 sources were included as targets for KVN observations 
(22 sources have CO(1$-$0) SD observations from literature).
Figure~\ref{fig:fluxcomp} displays the comparison between the 
CO(1$-$0) line luminosities from different observations. 
Among 63 sources with CO(1$-$0) SD observations, 
48 sources have multiple CO observations including other CO(1$-$0) SD, CO(1$-$0) array, 
CO(2$-$1) SD, CO(2$-$1) array,
and high-$J$ transition CO observations
(top panel of Figure~\ref{fig:fluxcomp}).
Using $\mathrm{log}\,L_\mathrm{CO(1-0)}^\prime$ value from one of the CO(1$-$0) SD observations 
(the value from the smallest telescope)
as a reference value (i.e., $\mathrm{log}\,L^\prime_\mathrm{CO(1-0),\,ref}$),
the mean difference 
($\Delta\,\mathrm{log}\,L^\prime_\mathrm{CO(1-0)}=\mathrm{log}\,L^\prime_\mathrm{CO(1-0)}-\mathrm{log}\,L^\prime_\mathrm{CO(1-0),\,ref}$)
among CO(1$-$0) SD observations is $-0.022\pm0.285$\,dex.
Including different transitions and observation types (SD or array),
the difference is $-0.054\pm0.362$\,dex. The intrinsic scatter 
in the $\mathrm{log}\,L_\mathrm{CO(1-0)}^\prime$ values is relatively large (i.e., $>0.3$\,dex).
Note that the inclusion of large sources (i.e., objects with an optical diameter larger than 100\arcsec)
tends to increase the scatter.
However, the systematic offset between the values from different observations 
does not change significantly by including such large sources, as well as sources that are located within the interacting/merging system.

\begin{figure*}
\epsscale{1.15}
\plottwo{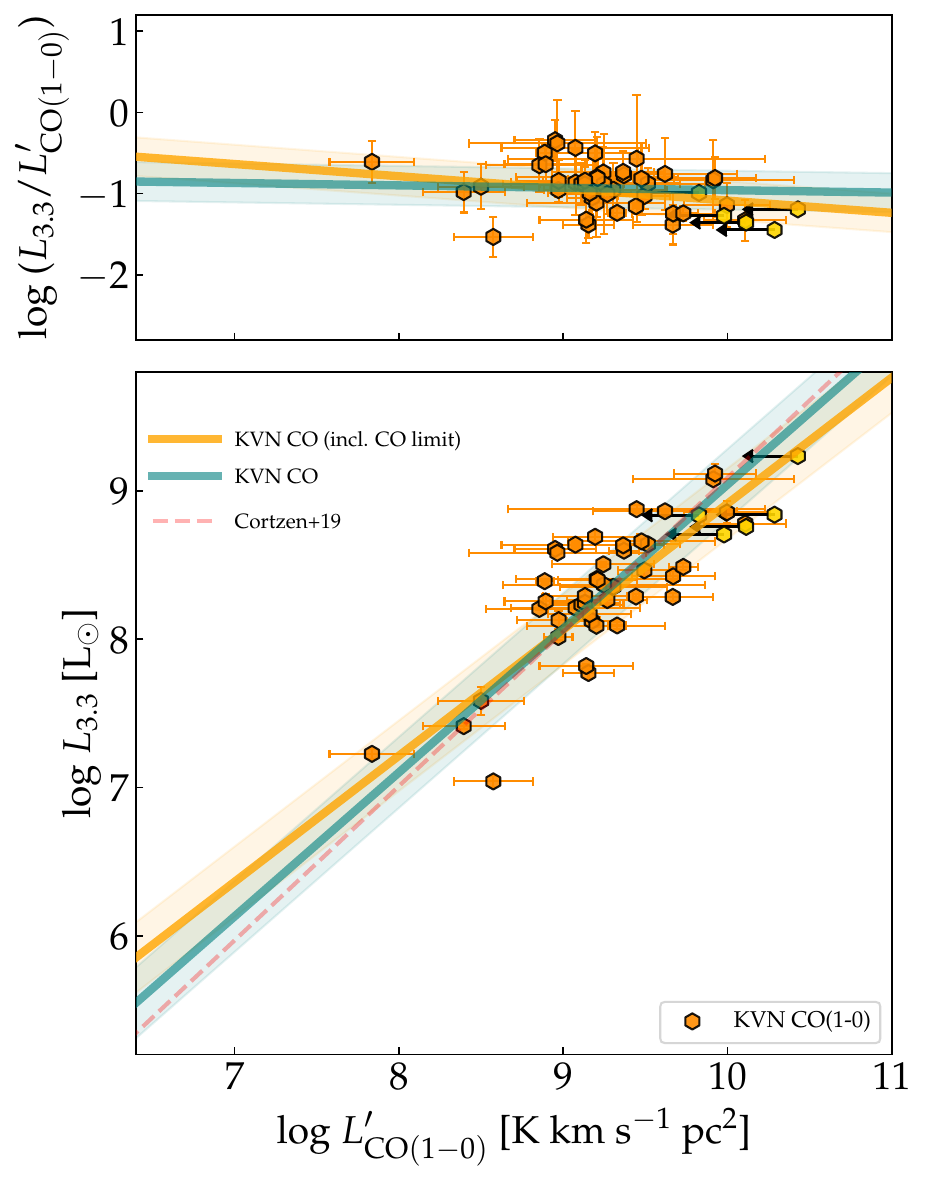}{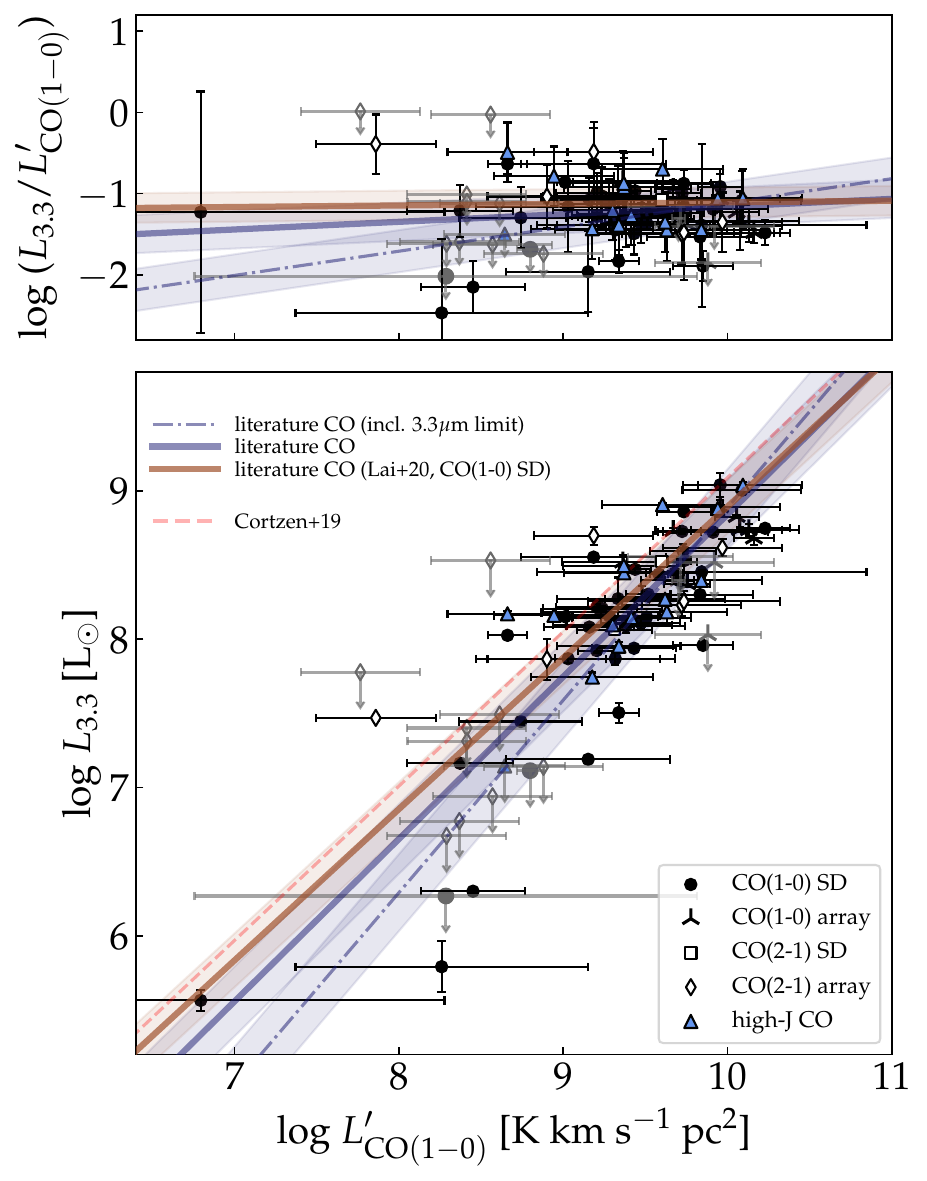}
\caption{\label{fig:corr}
Correlation between 3.3\,$\mu$m PAH luminosity and CO (1$-$0) luminosity 
constructed using KVN observations (Left) 
and literature data (Right, excluding KVN targets).
New KVN observations of 49 sources selected at 3.3\,$\mu$m
(excluded one source with 3.3\,$\mu$m upper limit)
are plotted as orange-colored hexagons (44), while five yellow hexagons with leftward arrows 
represent sources for which the CO line was not detected. 
Different symbols represent sources with 
literature-derived $\mathrm{log}\,L_\mathrm{CO}^\prime$ values: 
CO (1$-$0) single dish, CO (1$-$0) interferometer array, CO (2$-$1) single dish, 
CO (2$-$1) interferometer array, and higher-$J$ transitions.
Non-detections (i.e., upper limits) are plotted as gray colors. 
Different solid lines (and the dash-dot line) show the best-fit linear correlations between
$\mathrm{log}\,L_\mathrm{3.3}$ and $\mathrm{log}\,L_\mathrm{CO(1-0)}^\prime$
for different subsamples of sources.  
Shaded regions indicate the intrinsic scatter of the linear fits.
Dashed lines in the lower panels denote 
the 7.7\,$\mu$m PAH-CO relation 
suggested by \citet{2019MNRAS.482.1618C}, vertically shifted
by the average ratio between 3.3 and 7.7\,$\mu$m PAH emissions 
\citep{2020ApJ...905...55L}.
}
\end{figure*}

About half (26) of the KVN targets have CO measurements from literature
(bottom panel of Figure~\ref{fig:fluxcomp}). 
For these, the difference $\Delta\,\mathrm{log}\,L^\prime_\mathrm{CO(1-0)}$
is $0.226\pm0.239$\,dex, indicating that CO luminosities measured from KVN SD observations
are slightly lower compared to the values measured 
by other observations \citep[e.g., FCRAO;][]{1995ApJS...98..219Y}. 
One of the reasons could be a difference in the beam size 
between \citet{1995ApJS...98..219Y} 
(45\arcsec) and our work (30\arcsec) at $\sim115$\,GHz;
with the exclusion of \citet{1995ApJS...98..219Y} sample, $\Delta\,\mathrm{log}\,L^\prime_\mathrm{CO(1-0)}$
for the KVN targets is $0.165\pm0.230$. 
These differences in $\mathrm{log}\,L^\prime_\mathrm{CO(1-0)}$ 
were added to the uncertainties in CO luminosity. 
For KVN sources and literature sources with a single CO measurement, 
intrinsic scatters (i.e., 0.239 and 0.362) were added to uncertainties.
Considering the relatively large systematic offset of the KVN $L_\mathrm{CO}^\prime$ values
to that from previous literature, in the following section, 
we derived $\mathrm{log}\,L_{3.3}$-$\mathrm{log}\,L_\mathrm{CO(1-0)}^\prime$ relationships
based on the KVN observations and literature compilation separately,
in addition to using the entire combined sample.

\section{Results and Discussion}
\label{sec:results}

\begin{figure}
\epsscale{1.15}
\plotone{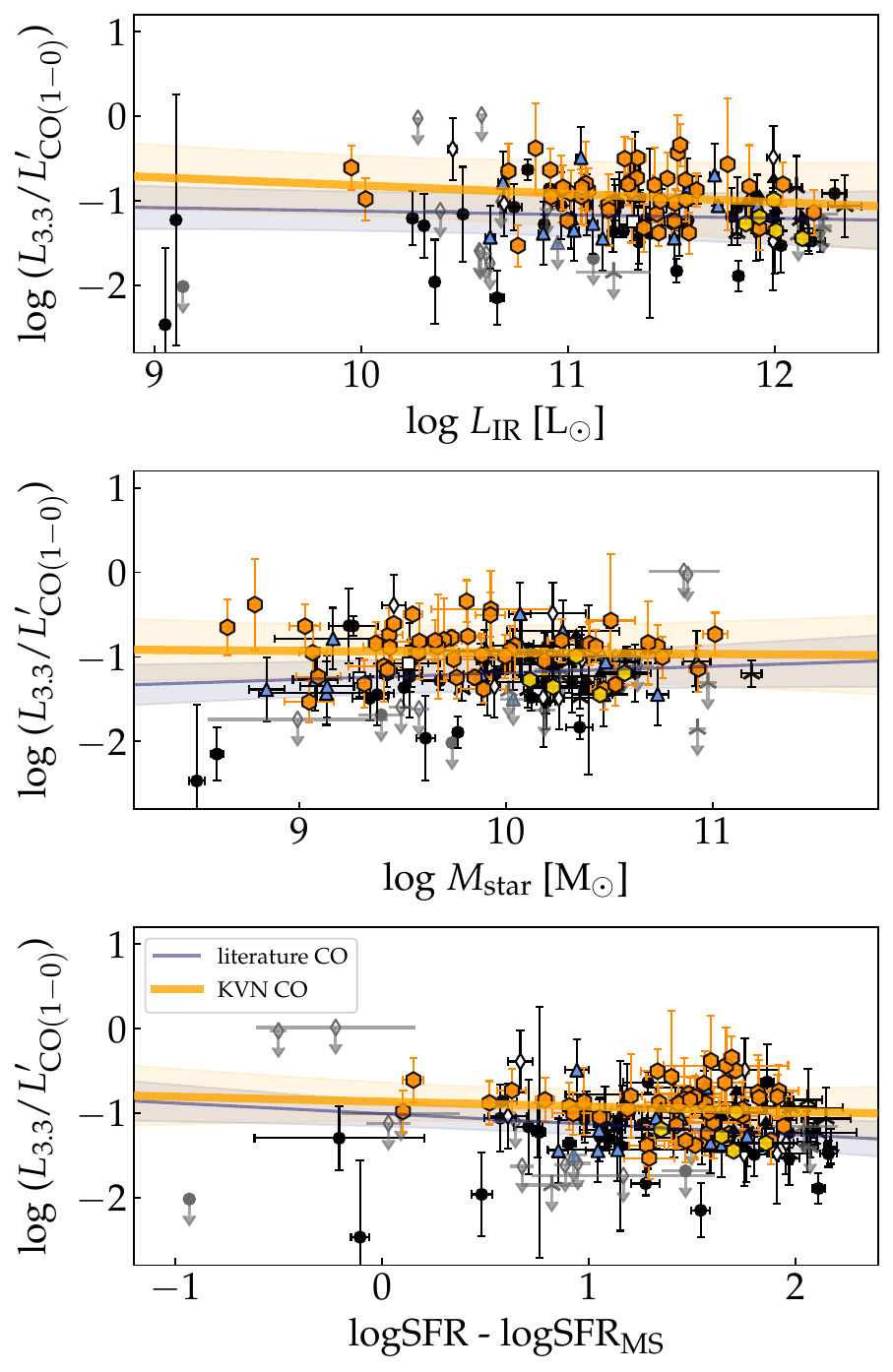}
\caption{ \label{fig:var33co}
The ratio between 3.3\,$\mu$m PAH and CO luminosities
in units of [L$_\odot$/(K\,km\,s$^{-1}$\,pc$^2$)]
as a function of total IR luminosity (top), stellar mass (middle), 
and offset from the star-forming galaxy main sequence (bottom). 
Symbols are as defined in Figure~\ref{fig:corr}.
Overplotted solid lines indicate linear correlations
constructed using CO luminosity data points compiled from the literature
and obtained from KVN observations. 
}
\end{figure}

\subsection{$L_\mathrm{3.3}$ vs. $L_\mathrm{CO(1-0)}^\prime$ correlation}    \label{sec:corr}

We show the correlation between the 3.3\,$\mu$m PAH luminosity 
and CO(1$-$0) luminosity in Figure~\ref{fig:corr}, 
using sources with our KVN observations 
as well as sources from the literature. 
The linear fit describing scaling relations between different luminosities 
is obtained in the logarithmic space,
with a form of $\mathrm{log}\,y=\alpha\times\mathrm{log}\,x+\beta$. 
The parameters, $\alpha$ and $\beta$, are summarized in Table~\ref{tab:fit}. 
The parameters are estimated by exploiting a Bayesian approach 
described in \citet{2007ApJ...665.1489K}, 
which derives a posterior distribution of the linear regression coefficients 
through an MCMC algorithm using a varying set of independent and dependent variables
using their errors (uncertainties) as dispersion of the distributions. 
The \textsc{python} implementation of the \citet{2007ApJ...665.1489K} method,
i.e., \texttt{LinMix}, allows the handling of non-detection (i.e., upper limits) 
by censoring specific data points.
We present the fitting results from different subsamples in Table~\ref{tab:fit} as well,
e.g., including CO non-detections (in the case of the KVN observations) 
and including 3.3\,$\mu$m upper limits (in the case of the literature compilation).

Using 49 KVN sources (3.3\,$\mu$m flux density available, 
including 5 sources with CO non-detection), 
the linear slope between $\mathrm{log}\,L_\mathrm{3.3}$ and $\mathrm{log}\,L_\mathrm{CO(1-0)}^\prime$
is slightly sub-unity ($0.85\pm0.12$).
Excluding CO non-detections, the slope is $0.97\pm0.14$, 
which is in line with the result of \citet{2019MNRAS.482.1618C}
stating that the $L_\mathrm{PAH}/L_\mathrm{CO}^\prime$ ratio is 
almost constant for star-forming galaxies in the case of 6.2\,$\mu$m and 7.7\,$\mu$m PAHs
(Figure~\ref{fig:corr}, left panel).
The mean $\mathrm{log}\,(L_\mathrm{3.3}/L_\mathrm{CO(1-0)}^\prime)$ value 
for 44 KVN sources is $-0.91$~$[\mathrm{L}_\odot/(\mathrm{K\,km\,s^{-1}\,pc^{2})}]$, 
with a standard deviation of $0.28$.
A similar relation is found from the linear fit for literature CO values; 
the slope is $1.09\pm0.11$ when sources with 3.3\,$\mu$m upper limits are not considered. 
As is discussed in Sections~\ref{sec:sample} and \ref{sec:compilation},
different treatment in dust attenuation correction 
and different CO observation strategies may affect the derived correlation between 
$L_{3.3}$ and $L_\mathrm{CO}^\prime$ using heterogeneous sample. 
Considering these, we fit the linear relation between 
$\mathrm{log}\,L_\mathrm{3.3}$ and $\mathrm{log}\,L_\mathrm{CO(1-0)}^\prime$ using a subsample 
of 49 sources from \citet{2020ApJ...905...55L} that have CO(1$-$0) SD measurements. 
For this subsample, the slope is $1.01\pm0.12$, 
consistent with the result from the entire literature sample and KVN sample 
with the exclusion of sources with upper limits (either in CO or 3.3\,$\mu$m).

If KVN and literature samples are combined (114 sources excluding upper limits), 
the relation between $\mathrm{log}\,L_\mathrm{3.3}/\mathrm{L}_\odot$ and
$\mathrm{log}\,L_\mathrm{CO(1-0)}^\prime/(\mathrm{K\,km\,s^{-1}\,pc^2})$
is described as follows:

\begin{equation}
   \mathrm{log}\,L_{3.3} = (1.00\pm0.07)\times\mathrm{log}\,L_\mathrm{CO(1-0)}^\prime+(-1.10\pm0.70).
   \label{eq:LCO33}
\end{equation}

The slope is close to unity ($1.00\pm0.07$), clearly suggesting
the possibility that PAH emission can be used to estimate 
the amount of molecular gas in galaxies.
The constant conversion factor between $L_{3.3}$ and $L_\mathrm{CO(1-0)}^\prime$ is:

\begin{equation}
    \mathrm{log}\,(L_{3.3}/L_\mathrm{CO(1-0)}^\prime)=-1.09\pm0.36,
\end{equation}

\noindent in units of $[\mathrm{L}_\odot/(\mathrm{K\,km\,s^{-1}\,pc^{2})}]$.



Note that the trend in $\mathrm{log}\,L_\mathrm{3.3}$-$\mathrm{log}\,L_\mathrm{CO(1-0)}^\prime$ space
remains unchanged if the large sources 
(optical diameter larger than 100\arcsec, marked in the top panel of Figure~\ref{fig:fluxcomp}) 
and/or interacting/merging galaxies
are excluded.
To explore factors responsible for the scatter 
in the observed $L_\mathrm{3.3}$-$L_\mathrm{CO}^\prime$ correlation, 
we display the $L_\mathrm{3.3}/L_\mathrm{CO}^\prime$ ratios 
as a function of IR luminosity, stellar mass, and star formation activity 
(Figure~\ref{fig:var33co}). 
Star formation activity is defined as the offset from the star-forming main sequence,
where $\mathrm{SFR}_\mathrm{MS}$ is a function of stellar mass and redshift
\citep{2014ApJS..214...15S}. 
In all three panels, 
the Spearman coefficients are too low (with $p$-values exceeding 0.1),
thus it cannot be concluded that the scatter in 3.3\,$\mu$m PAH-CO ratios
is driven by the different physical properties of galaxies.

\subsection{Molecular gas mass and $L_\mathrm{3.3}$}  \label{sec:gasmass}

The molecular gas mass of a galaxy can be estimated 
from the observed CO luminosity by applying a CO-$\mathrm{H}_2$ conversion factor
(i.e., $M(\mathrm{H}_2)\equiv\alpha_\mathrm{CO} \times L_\mathrm{CO}^\prime$).
The $\alpha_\mathrm{CO}$ factor 
exhibits large variation for different galaxies 
\citep[e.g.,][]{2011ApJ...737...12L, 2012ApJ...751...10P},
and is suggested to be governed by interstellar medium characteristics in 
star forming regions such as the metallicity, gas temperature, velocity dispersion,
and surface density
\citep{2012MNRAS.421.3127N, 2013ARA&A..51..207B}. 
The $\alpha_\mathrm{CO}$ factor for 
normal main sequence star-forming galaxies
is similar to the average value in Milky Way disks 
\citep{1987ApJ...319..730S}, i.e.,
$\langle\alpha_\mathrm{CO}\rangle\sim4.4\,\mathrm{M}_\odot$/(K km s$^{-1}$ pc$^2$),
while the $\alpha_\mathrm{CO}$ factor 
for extreme starburst environments is lower
\citep[$\sim0.8$;][]{1998ApJ...507..615D} 
when measured for local ULIRGs and high-redshift starburst galaxies, 
due to the increased gas temperature and velocity dispersion. 
In a low-metallicity environment with less metal shielding, however, 
$\alpha_\mathrm{CO}$ is reported to increase \citep{2012MNRAS.421.3127N},
as CO molecules are destroyed through photo-dissociation. 
To overcome the issues related to this $\alpha_\mathrm{CO}$ uncertainty, 
alternative methods, such as using dust emission
and scaling relations to estimate gas mass, 
are being introduced and utilized,
especially in surveys with (rest-frame) far-infrared 
photometric information \citep[e.g.,][]{2016A&A...587A..73B, 2018MNRAS.478.1442B}.

\begin{figure}
\epsscale{1.15}
\plotone{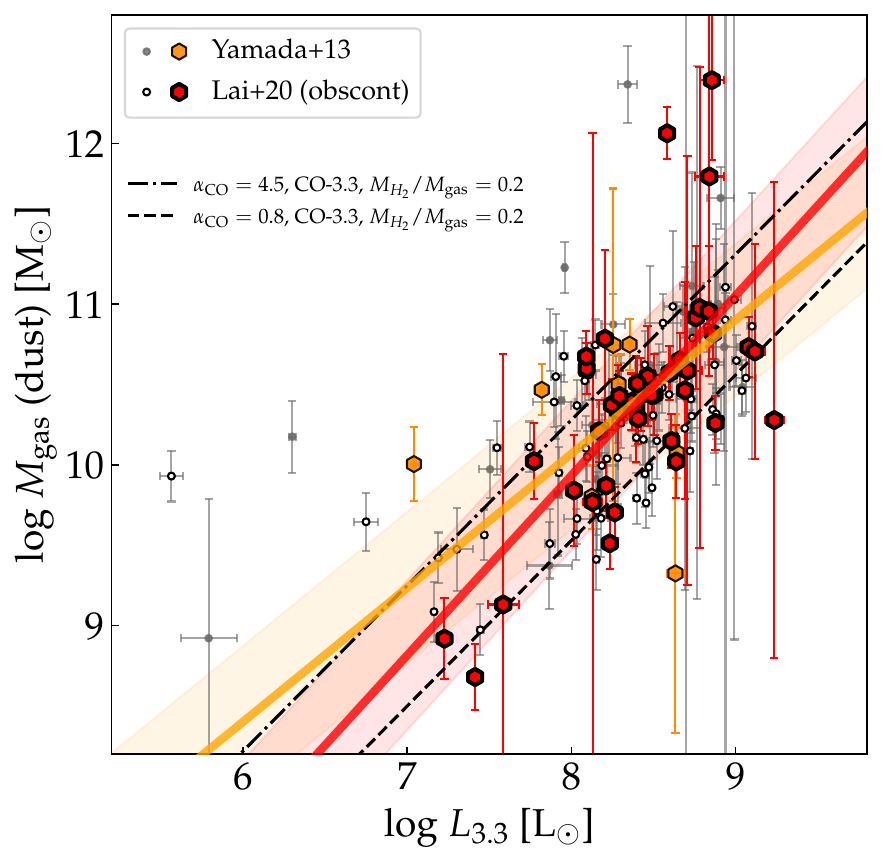}
\caption{ \label{fig:mgas}
Correlation between the gas mass estimated from dust mass
($M_\mathrm{gas}\equiv M_\textsc{hi}+M_\mathrm{H_2}$) 
and 3.3\,$\mu$m PAH luminosity. 
Colored symbols represent KVN targets (including CO non-detections), 
while the white/gray symbols are sources from the literature.
Dash-dot and dashed lines 
represent the $M_\mathrm{H_2}$ vs. CO luminosity relationships assuming
$M_\mathrm{H_2}/M_\mathrm{gas}=0.2$ in addition to
$\alpha_\mathrm{CO}=4.5$ and 0.8, respectively, 
of which CO luminosity is converted from the 3.3\,$\mu$m PAH luminosity
using the derived correlation in Figure~\ref{fig:corr} (teal line using
44 KVN sources; Table~\ref{tab:fit}).
Overplotted solid lines are the best-fit linear relationships between
$\mathrm{log}\,M_\mathrm{gas}$ and $\mathrm{log}\,L_{3.3}$
using all 49 KVN sources (orange and red) and 
41 KVN sources from \citet{2020ApJ...905...55L} (red).}
\end{figure}

\begin{figure*}
\epsscale{1.15}
    \plottwo{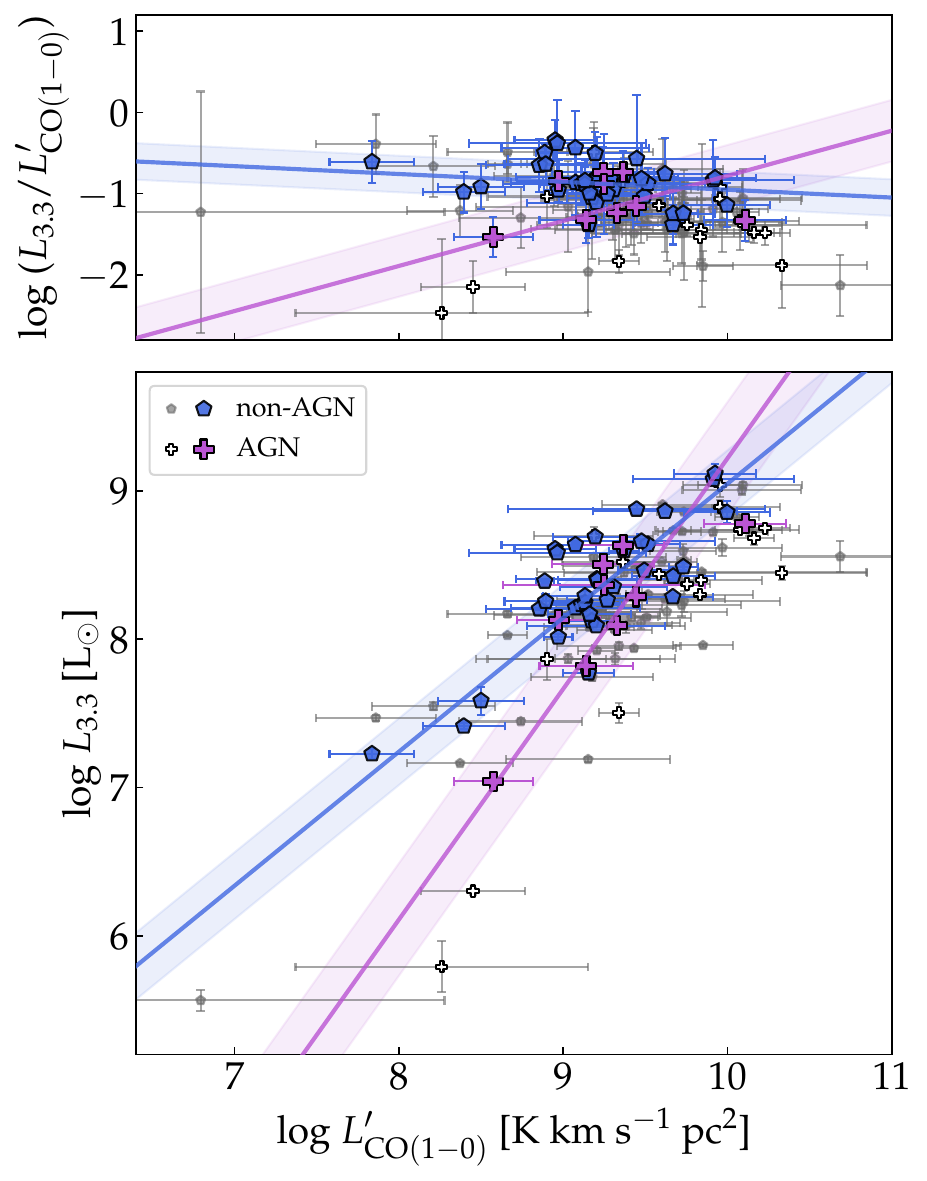}{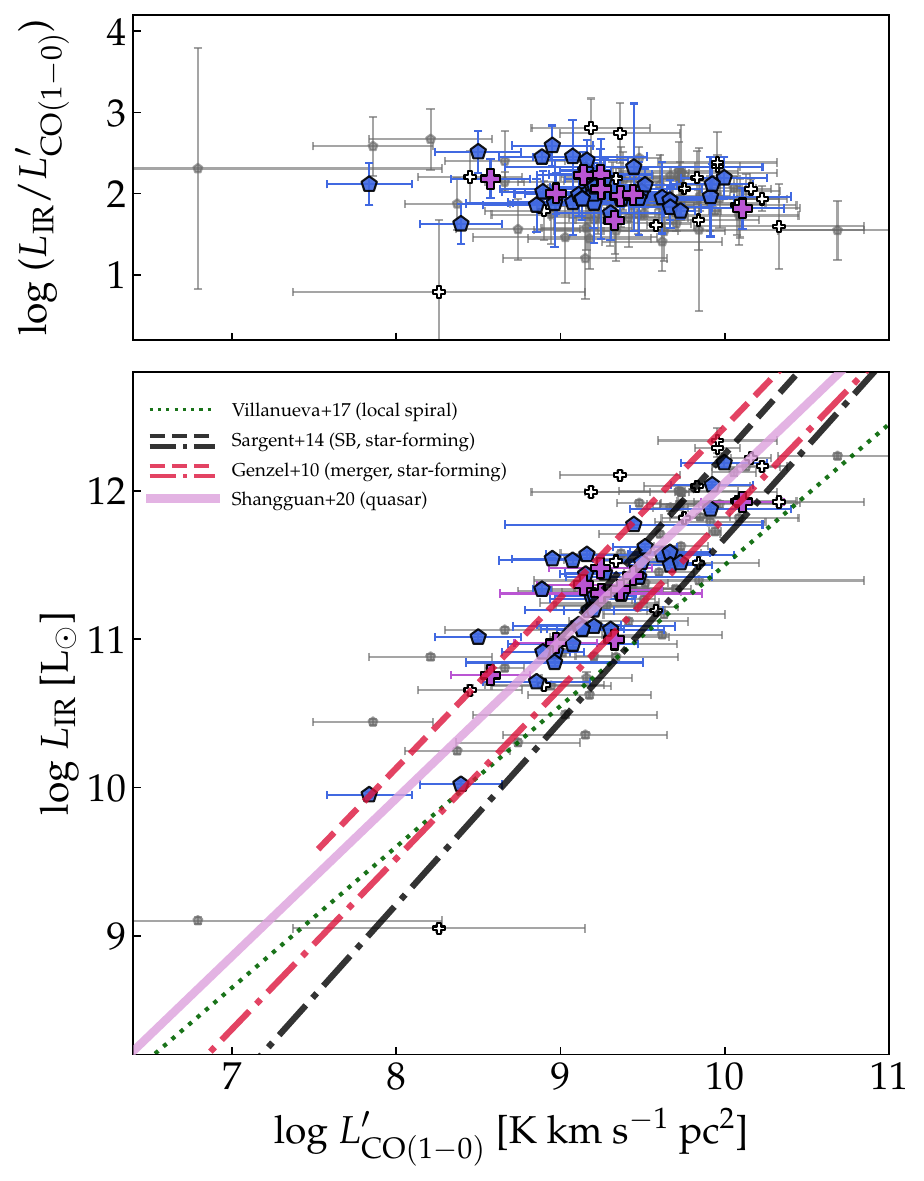}
    \caption{\label{fig:agn} 
    (Left) Correlation between 3.3\,$\mu$m PAH luminosity and CO luminosity 
    for AGN (crosses) and non-AGN sources (pentagons). 
    Colored symbols (purple crosses, blue pentagons) represent KVN sources, 
    while the white/grey symbols represent sources from the literature. 
    Best-fit linear relation between $\mathrm{log}\,L_{3.3}$ and $\mathrm{log}\,L_\mathrm{CO(1-0)}^\prime$
    are constructed using the KVN sources, excluding non-detections.
    (Right) Correlation between total IR luminosity and CO luminosity
    for AGN and non-AGN. Overplotted lines represent the 
    $\mathrm{log}\,L_\mathrm{IR}$-$\mathrm{log}\,L_\mathrm{CO(1-0)}^\prime$ calibrations 
    from different literature; 
    green dotted line for local spiral galaxies \citep{2017MNRAS.470.3775V},
    dash-dot lines for star-forming galaxies (black from \citet{2014ApJ...793...19S} and crimson from
    \citet{2010MNRAS.407.2091G}, respectively),
    dashed lines for starbursts \citep{2014ApJ...793...19S} and luminous mergers \citep{2010MNRAS.407.2091G},
    and purple solid line for quasar hosts \citep{2020ApJS..247...15S}.
    For starbursts, the lines are drawn in the IR luminosity range that cover the sample galaxies
    used in deriving the calibrations.
    }
\end{figure*}

For 3.3\,$\mu$m-selected sources with KVN CO(1$-$0) observations, 
we estimated their gas mass ($M_\mathrm{gas}$)
based on the dust mass ($M_\mathrm{dust}$) and gas-to-dust mass ratio 
($\delta_\mathrm{GDR}\equiv M_\mathrm{gas}/M_\mathrm{dust}$). 
The dust mass of the sources was estimated by fitting IR SED to dust emission models 
\citep{2014ApJ...780..172D} within the SED fitting code \textsc{cigale} \citep{2019A&A...622A.103B}
as described in Section~\ref{sec:sample},
assuming that the dust emission is proportional to the dust mass 
illuminated by a power-law shaped radiation field intensity. 
We used the metallicity-dependent gas-to-dust mass ratio 
\citep{2012ApJ...760....6M}
by applying the (gas-phase) metallicity inferred through 
the fundamental mass-SFR-metallicity relation \citep{2010MNRAS.408.2115M} 
along with the SED fitting-derived stellar mass and SFR for these sources. 
Note that the uncertainty in the derived gas mass is generally large, 
reflecting photometric uncertainties in the FIR wavelengths, 
uncertainties in the derived stellar mass and SFR used to estimate metallicity,
and uncertainties inherent in the empirical relations as well.  

Figure~\ref{fig:mgas} shows the correlation between 
the gas mass estimated from the dust emission 
and 3.3\,$\mu$m luminosity. 
The gas mass described here is the sum of the molecular hydrogen 
and atomic hydrogen, 
and the mass ratio $M_\mathrm{H_2}/M_\mathrm{\textsc{hi}}$ varies 
across different star-forming environments, 
with a connection to the surface density.
This complicates the interpretation of the $M_\mathrm{gas}$-$L_\mathrm{3.3}$ correlation. 
Despite a large scatter, the Spearman rank coefficient between 
$\mathrm{log}\,M_\mathrm{gas}$ and $\mathrm{log}\,L_\mathrm{3.3}$ is 
statistically significant (with a $p$-value less than $10^{-4}$), 
and the correlation is approximated by the following
linear relation with a scatter of $\sigma\sim0.46$:

\begin{equation}
    \mathrm{log}\,M_\mathrm{gas} = (0.83\pm0.19)\times\mathrm{log}\,L_\mathrm{3.3}+(3.40\pm1.63). 
\end{equation}

If the sample is limited to 41 sources with 3.3\,$\mu$m measurements 
from \citet{2020ApJ...905...55L}, 
the correlation is slightly different, and described as follows:

\begin{equation}
    \mathrm{log}\,M_\mathrm{gas} = (1.12\pm0.23)\times\mathrm{log}\,L_\mathrm{3.3}+(0.98\pm1.92). 
\end{equation}

The $1\sigma$ uncertainty regions suggested by these correlations 
are comparable to the space covered by 
$M_\mathrm{H_2}=\alpha_\mathrm{CO}L_\mathrm{CO}^\prime$ correlation 
with $\alpha_\mathrm{CO}=0.8$-4.5, 
assuming a mass ratio $M_\mathrm{H_2}/M_\mathrm{gas}$
of 0.2 \citep[for late-type spirals;][]{1991ARA&A..29..581Y}. 
The range of the assumed $\alpha_\mathrm{CO}$ values
suggests that the derived $\mathrm{log}\,M_\mathrm{gas}$-$\mathrm{log}\,L_{3.3}$ correlation
is applicable to the wide range of star-forming galaxies, from normal disk galaxies to extreme starbursts.


Although our KVN targets did not include sources with $\mathrm{log}\,(L_{3.3}/\mathrm{L_\odot})<7$,
Figure~\ref{fig:mgas} shows that 
there exist several sources that are expected to have significant amounts of gas
despite low luminosity,
e.g., low-metallicity galaxy II\,Zw\,40 with $\mathrm{log}\,(L_{3.3}/\mathrm{L_\odot})<6$. 
The presence of these increases scatter in the derived linear relation between
$\mathrm{log}\,M_\mathrm{gas}$ and $\mathrm{log}\,L_{3.3}$,
and needs to be explored further with the increased number of sources
for which $L_{3.3}$ is measured at a low luminosity range.

\subsection{$L_\mathrm{3.3}$, $L_\mathrm{IR}$, and $L_\mathrm{CO}^\prime$ for AGN}     \label{sec:agn}

As a star formation tracer, PAH emission has been used to estimate 
the SFR in AGN host galaxies \citep{2008ApJ...684..853L, 2019PASJ...71...25K}
as well as in star-forming galaxies.
Since the radiation field characteristics in the vicinity of an AGN 
(e.g., close to the central region of an AGN host galaxy) 
and in the average star-forming disk could differ, 
the PAH-SFR calibration for AGN might be different from that generally
used for star-forming galaxies \citep{2004ApJ...613..986P}.
PAH molecule destruction in the strong radiation field \citep{1992MNRAS.258..841V, 2007ApJ...669..810D}
may lead to a weak PAH emission in the spectra of AGN host galaxies. 

In Figure~\ref{fig:agn}, we present $L_\mathrm{3.3}$-$L_\mathrm{CO}^\prime$ correlations 
and $L_\mathrm{IR}$-$L_\mathrm{CO}^\prime$ correlations for
AGN and non-AGN (i.e., star-forming galaxies) separately (listed in Table~\ref{tab:fit}). 
The two populations are not distinguished in the
$\mathrm{log}\,L_\mathrm{IR}$ vs. $\mathrm{log}\,L_\mathrm{CO}^\prime$ space (right panel).
However, in the $\mathrm{log}\,L_\mathrm{3.3}$ vs. $\mathrm{log}\,L_\mathrm{CO}^\prime$ diagram (left panel), 
the two populations are discriminated;
the slope in the $\mathrm{log}\,L_\mathrm{3.3}$-$\mathrm{log}\,L_\mathrm{CO}^\prime$ 
relation is steeper in the case of the AGN ($1.56\pm0.84$) than in the case of non-AGN ($0.90\pm0.15$).
This implies that in AGN with low CO luminosity, 
3.3\,$\mu$m PAH emission is weaker than in star-forming galaxies.

Previous works have suggested the scatter in $\mathrm{log}\,L_\mathrm{IR}$-$\mathrm{log}\,L_\mathrm{CO}^\prime$
can be either explained as a separate $\mathrm{log}\,(L_\mathrm{IR}/L_\mathrm{CO(1-0)}^\prime)$ ratio 
for starburst and normal star-forming galaxies (with the latter being lower than the former),
or a single relation $L_\mathrm{IR}\propto {L_\mathrm{CO}^\prime}^{k}$ where the slope $k$ is larger than unity 
\citep{2014ApJ...793...19S}. 
In the right panel of Figure~\ref{fig:agn}, 
the KVN sources are located in between the $\mathrm{log}\,L_\mathrm{IR}$-$\mathrm{log}\,L_\mathrm{CO}^\prime$ trends
of the star-forming and starburst galaxies \citep[e.g.,][]{2010MNRAS.407.2091G, 2014ApJ...793...19S},
and are also consistent with that of the quasar host galaxies \citep{2020ApJS..247...15S}. 
Although the number of AGN in our KVN sample is very small (only nine, excluding non-detections),
the fact that the AGN show comparable slope in 
$\mathrm{log}\,L_\mathrm{IR}$ vs. $\mathrm{log}\,L_\mathrm{CO}^\prime$ space 
suggests that the global star formation efficiency in AGN host galaxies 
is not significantly enhanced \citep{2012A&A...540A.109S} or suppressed 
compared to inactive star-forming galaxies.

However, it is still a matter of question whether 3.3\,$\mu$m PAH emission 
can serve as a reliable tracer of SFR in AGN host galaxies, 
as in the case of the other PAH features at longer wavelengths. 
Further studies on the ratios between different PAH feature strengths 
\citep[e.g.,][]{2021MNRAS.504.5287R} using the spatially resolved data \citep[e.g.,][]{2023ApJ...957L..26L}
and the increased sample size from upcoming near-infrared spectrophotometric surveys
would be required to assess the viability of using 3.3\,$\mu$m PAH as an SFR indicator.

\section{Conclusion}

Combining new CO(1$-$0) observations of 50 3.3$\mu$m-selected sources (including both star-forming galaxies and AGN)
with an existing literature compilation of CO data,
we have constructed a correlation between 3.3\,$\mu$m PAH luminosity 
and CO luminosity. This correlation is applicable to estimate
(molecular) gas mass and star formation efficiency in galaxies. 
Our sample galaxies cover an IR luminosity range of 
$\mathrm{log}\,(L_\mathrm{IR}/\mathrm{L}_\odot)=10$-12.
The sample size of galaxies for which both 3.3\,$\mu$m and CO(1$-$0) observations are 
available has been increased by a factor of $\sim1.6$ by our CO observations.

In the $\mathrm{log}\,L_\mathrm{3.3}$-$\mathrm{log}\,L_\mathrm{CO}^\prime$ space, 
we establish a linear correlation between the two variables,
using $\mathrm{log}\,L_\mathrm{CO}^\prime$ as the independent variable. 
The slope is close to unity, 
suggesting that it is possible to use a single value of 
$\langle \mathrm{log}\,(L_\mathrm{3.3}/L_\mathrm{CO}^\prime)\rangle=-1.09$ 
for $L_\mathrm{3.3}$ in units of $\mathrm{L}_\odot$ and 
$L_\mathrm{CO}^\prime$ in units of $\mathrm{K\,km\,s^{-1}\,pc^2}$,
with a scatter of $\sim0.36$ (based on the entire sample including KVN and literature, 
excluding non-detections).
The scatter in the $L_\mathrm{3.3}/L_\mathrm{CO}^\prime$ 
is not correlated with IR luminosity, stellar mass, and SFR excess. 
Based on the derived conversion formula between $\mathrm{log}\,L_\mathrm{3.3}$
and $\mathrm{log}\,L_\mathrm{CO}^\prime$, we suggest a scaling relation between 
the 3.3\,$\mu$m luminosity and gas mass (total gas mass, defined as the sum of molecular and atomic mass),
which overlaps with previous CO-$\mathrm{H_2}$ conversion factors 
for starbursts to main-sequence star-forming galaxies.
The global star formation efficiency, reflected in the $L_\mathrm{IR}/L_\mathrm{CO}^\prime$,
for AGN and non-AGN, is consistent with each other based on the KVN sources. 
However, it is suspected that AGN show different trends 
in the $\mathrm{log}\,L_\mathrm{3.3}$-$\mathrm{log}\,L_\mathrm{CO}^\prime$ space
compared to non-AGN, which needs to be verified by further studies using the increased number of 3.3\,$\mu$m-selected sources.
With the upcoming availability of all-sky spectrophotometric surveys 
covering wavelengths 1-5\,$\mu$m, 
the potential of using 3.3\,$\mu$m PAH emission in estimating gas mass
would be tested further, for which scaling relation can be applied to galaxies 
at various cosmic epochs.

\startlongtable
\tabletypesize{\footnotesize}
\begin{deluxetable*}{lcccccccc}
\def\arraystretch{1}
\tablecaption{\label{tab:summary} Summary of the KVN SD CO(1$-$0) observations of the 3.3\,$\mu$m-PAH-selected galaxies}
\tablehead{\colhead{Name} & \colhead{R.A.} & \colhead{Decl.} & \colhead{$z_\mathrm{CO}$} & \colhead{$S_\mathrm{CO} \Delta v$} & \colhead{FWHM} & \colhead{log\,$L_\mathrm{3.3}$} & \colhead{log\,$L_\mathrm{CO(1-0)}^\prime$} & \colhead{rms} \\
\colhead{} & \colhead{[hh:mm:ss]} & \colhead{[dd:mm:ss]} & \colhead{} & \colhead{[Jy km s$^{-1}$]} & \colhead{[km s$^{-1}$]} & \colhead{[$\mathrm{L}_\odot$]} & \colhead{[K km s$^{-1}$ pc$^2$]} & \colhead{[mK]} \\
\colhead{(1)} & \colhead{(2)} & \colhead{(3)} & \colhead{(4)} & \colhead{(5)} & \colhead{(6)} & \colhead{(7)} & \colhead{(8)} & \colhead{(9)}  
}
    \startdata
    NGC 0023      & 00:09:53.41 & +25:55:25.6 & 0.0153 & $113.3\pm10.5$ & 352 & $8.21\pm0.01$ & $9.08\pm0.40$ & 2.1  \\
    MCG+12-02-001\tablenotemark{\dag} & 00:54:03.61 & +73:05:11.8 & 0.0162 & $69.1\pm6.5$ & 144 & $8.39\pm0.01$ & $8.89\pm0.06$ & 2.9  \\
    CGCG 436-030  & 01:20:02.72 & +14:21:42.9 & 0.0313 & $39.4\pm6.1$ & 209 & $8.51\pm0.02$ & $9.24\pm0.31$ & 2.2  \\
    III Zw 035    & 01:44:30.50 & +17:06:05.0 & 0.0275 & $40.7\pm4.4$ & 222 & $7.77\pm0.04$ & $9.15\pm0.16$ & 1.4  \\
    NGC 0695      & 01:51:14.24 & +22:34:56.5 & 0.0324 & $86.5\pm10.6$ & 211 & $8.86\pm0.01$ & $9.62\pm0.44$ & 3.2  \\
    UGC 01385     & 01:54:53.79 & +36:55:04.6 & 0.0185 & $58.6\pm5.5$ & 95 & $8.02\pm0.02$ & $8.97\pm0.09$ & 3.4  \\
    UGC 01845     & 02:24:07.98 & +47:58:11.0 & 0.0147 & $117.2\pm16.5$ & 281 & $8.23\pm0.02$ & $9.11\pm0.26$ & 5.1  \\
    NGC 0992      & 02:37:25.49 & +21:06:03.0 & 0.0136 & $83.0\pm8.2$ & 283 & $8.20\pm0.01$ & $8.85\pm0.33$ & 2.3  \\
    UGC 02238     & 02:46:17.49 & +13:05:44.4 & 0.0219 & $107.8\pm10.5$ & 347 & $8.60\pm0.01$ & $9.37\pm0.10$ & 2.9  \\
    UGC 02369\tablenotemark{\dag}  & 02:54:01.78 & +14:58:25.0 & 0.0316 & $70.2\pm10.1$ & 255 & $8.46\pm0.04$ & $9.49\pm0.16$ & 2.5  \\
    SDSS J032322.86-075615.2 & 03:23:22.86 & -07:56:15.3 & 0.1664\tablenotemark{\ddag} & $<14.4$ & - & $8.84\pm0.09$ & $<10.29$ & 0.8  \\
    UGC 02982     & 04:12:22.45 & +05:32:50.6 & 0.0178 & $122.2\pm7.5$ & 362 & $8.41\pm0.01$ & $9.20\pm0.50$ & 1.8  \\
    NGC 1614      & 04:33:59.85 & -08:34:43.9 & 0.0159 & $103.2\pm9.1$ & 245 & $8.64\pm0.01$ & $9.07\pm0.45$ & 2.3  \\
    VII Zw 031    & 05:16:46.09 & +79:40:13.2 & 0.0544 & $61.7\pm7.9$ & 169 & $9.08\pm0.01$ & $9.91\pm0.49$ & 2.8   \\
    NGC 2388      & 07:28:53.44 & +33:49:08.7 & 0.0136 & $173.5\pm9.6$ & 253 & $8.12\pm0.01$ & $9.17\pm0.15$ & 2.2  \\
    2MASX J08310928+5533165 & 08:31:09.29 & +55:33:16.4 & 0.0448 & $3.4\pm0.9$ & 51 & $7.58\pm0.07$ & $8.50\pm0.26$ & 1.2  \\
    NGC 2623      & 08:38:24.07 & +25:45:16.7 & 0.0184 & $88.9\pm9.2$ & 268 & $7.82\pm0.01$ & $9.14\pm0.28$ & 1.9  \\
    UGC 04730     & 09:01:58.39 & +60:09:06.1 & 0.0113 & $44.6\pm7.7$ & 735 & $7.41\pm0.02$ & $8.39\pm0.25$ & 1.0  \\
    2MASX J09024894+5236247 & 09:02:48.90 & +52:36:24.6 & 0.1573 & $8.3\pm2.0$ & 333 & $8.86\pm0.07$ & $10.00\pm0.26$ & 0.5  \\
    UGC 04881\tablenotemark{\dag}   & 09:15:55.10 & +44:19:55.0 & 0.0399 & $65.6\pm6.4$ & 247 & $8.29\pm0.03$ & $9.67\pm0.24$ & 1.8  \\
    NGC 3110      & 10:04:02.11 & -06:28:29.2 & 0.0169 & $157.0\pm11.3$ & 303 & $8.35\pm0.01$ & $9.31\pm0.33$ & 2.9  \\
    2MASX J10522356+4408474 & 10:52:23.52 & +44:08:47.7 & 0.0921\tablenotemark{\ddag} & $<32.6$ & - & $8.76\pm0.04$ & $<10.11$ & 2.1  \\
    Arp 148\tablenotemark{\dag}    & 11:03:53.20 & +40:50:57.0 & 0.0348 & $85.8\pm16.3$ & 487 & $8.42\pm0.03$ & $9.67\pm0.25$ & 2.4  \\
    SBS 1133+572  & 11:35:49.07 & +56:57:08.2 & 0.0517 & $19.6\pm3.4$ & 518 & $8.63\pm0.05$ & $9.37\pm0.25$ & 0.6  \\
    Mrk 1457      & 11:47:21.61 & +52:26:58.5 & 0.0487 & $8.6\pm1.6$ & 126 & $8.13\pm0.06$ & $8.97\pm0.25$ & 0.8  \\
    IC 0836       & 12:55:54.02 & +63:36:44.4 & 0.0093 & $17.9\pm4.1$ & 360 & $7.23\pm0.04$ & $7.84\pm0.26$ & 1.6  \\
    VV 283a       & 13:01:50.26 & +04:20:01.9 & 0.0375 & $83.7\pm12.7$ & 359 & $8.49\pm0.02$ & $9.73\pm0.09$ & 2.7  \\
    IC 0860       & 13:15:03.52 & +24:37:07.9 & 0.0130 & $50.0\pm3.6$ & 183 & $7.04\pm0.04$ & $8.57\pm0.24$ & 1.2  \\
    UGC 08335\tablenotemark{\dag}  & 13:15:32.80 & +62:07:37.0 & 0.0315 & $21.5\pm3.2$ & 275 & $8.61\pm0.04$ & $8.95\pm0.25$ & 0.9  \\
    NGC 5104      & 13:21:23.09 & +00:20:32.6 & 0.0185 & $135.7\pm17.1$ & 488 & $8.209\pm0.02$ & $9.33\pm0.06$ & 2.7  \\
    NGC 5256\tablenotemark{\dag}   & 13:38:17.50 & +48:16:37.0 & 0.0283 & $51.0\pm5.1$ & 343 & $8.37\pm0.02$ & $9.25\pm0.62$ & 1.0  \\
    2MASX J13561001+2905355 & 13:56:10.00 & +29:05:35.2 & 0.1089 & $23.1\pm4.0$ & 494 & $8.78\pm0.03$ & $10.11\pm0.25$ & 1.1  \\
    UGC 09618\tablenotemark{\dag}   & 14:57:00.40 & +24:36:44.0 & 0.0337 & $17.8\pm1.3$ & 545 & $8.58\pm0.01$ & $8.96\pm0.54$ & 1.0  \\
    SDSS J150539.56+574307.2 & 15:05:39.56 & +57:43:07.3 & 0.1506\tablenotemark{\ddag} & $<8.74$ & - & $8.71\pm0.09$ & $<9.98$ & 0.6  \\
    VV 705        & 15:18:06.34 & +42:44:36.6 & 0.0405 & $44.3\pm4.8$ & 176 & $8.64\pm0.01$ & $9.52\pm0.19$ & 1.5  \\
    NGC 5992      & 15:44:21.51 & +41:05:10.9 & 0.0320 & $17.0\pm3.0$ & 175 & $8.26\pm0.05$ & $8.89\pm0.25$ & 1.3  \\
    NGC 6090      & 16:11:40.70 & +52:27:24.0 & 0.0295 & $71.9\pm4.5$ & 118 & $8.66\pm0.02$ & $9.48\pm0.45$ & 2.0  \\
    SDSS J163452.36+462452.5 & 16:34:52.36 & +46:24:52.5 & 0.1908\tablenotemark{\ddag} & $<15.1$ & - & $9.24\pm0.06$ & $<10.43$ & 0.8  \\
    2MASX J16491420+3425096  & 16:49:14.20 & +34:25:09.7 & 0.1125 & $14.3\pm2.5$ & 678 & $9.12\pm0.06$ & $9.92\pm0.25$ & 0.7  \\
    CGCG 367-020  & 16:52:36.82 & +81:00:16.6 & 0.0472 & $14.1\pm3.2$ & 160 & $8.69\pm0.04$ & $9.19\pm0.26$ & 1.7  \\
    NGC 6286      & 16:58:31.38 & +58:56:10.5 & 0.0189 & $104.3\pm18.6$ & 560 & $8.09\pm0.01$ & $9.20\pm0.42$ & 3.4  \\
    2MASX J17034196+5813443  & 17:03:41.94 & +58:13:44.7 & 0.1061\tablenotemark{\ddag} & $<12.6$ & - & $8.84\pm0.05$ & $<9.83$ & 0.8  \\
    IRAS 17132+5313  & 17:14:20.00 & +53:10:30.0 & 0.0513 & $23.4\pm5.8$ & 465 & $8.88\pm0.03$ & $9.45\pm0.78$ & 1.4  \\
    NGC 6670A     & 18:33:37.72 & +59:53:22.8 & 0.0291 & $42.0\pm2.4$ & 149 & $8.40\pm0.05$ & $9.21\pm0.24$ & 1.1  \\
    NGC 6786      & 19:10:53.91 & +73:24:36.6 & 0.0252 & $47.6\pm3.5$ & 172 & $8.25\pm0.01$ & $9.13\pm0.21$ & 1.2  \\
    NVSS J211129+582307  & 21:11:29.28 & +58:23:07.9 & 0.0393 & $21.7\pm4.0$ & 255 & $8.17\pm0.05$ & $9.16\pm0.25$ & 1.4  \\
    3C 445        & 22:23:49.53 & -02:06:12.9 & 0.0559\tablenotemark{\ddag} & $<12.0$ & - & $<8.26$ & $<9.24$ & 0.6  \\
    NGC 7469      & 23:03:15.62 & +08:52:26.3 & 0.0164 & $230.0\pm4.0$ & 208 & $8.29\pm0.02$ & $9.44\pm0.02$ & 1.2  \\
    Arp 295B      & 23:42:00.85 & -03:36:54.6 & 0.0231 & $55.3\pm10.7$ & 411 & $8.29\pm0.04$ & $9.13\pm0.25$ & 1.9  \\
    2MASX J23460571+5313564  & 23:46:05.58 & +53:14:00.6 & 0.0340 & $34.9\pm3.7$ & 186 & $8.26\pm0.04$ & $9.27\pm0.24$ & 1.2  \\
    \enddata
    \tablecomments{Column (1): name of the source. Columns (2)-(3): coordinate of the source, defined as the KVN pointing coordinate. Column (4): redshift measured from the shift of CO(1$-$0) line, accounting for the velocity offset. Column (5): velocity-integrated flux within the integral interval (gray regions in Figure~\ref{fig:spec}). Column (6): FWHM measured using single Gaussian fit or line width in the horn-shaped profile. Column (7): $\mathrm{log}\,L_{3.3}$ compiled from literature. Column (8): $\mathrm{log}\,L_\mathrm{CO(1-0)}^\prime$ calculated from the KVN observations. Column (9): rms in the $T_\mathrm{A}^{*}$ in the KVN observations.}
\tablenotetext{\dag}{Merger/interacting source for which pointing coordinates for 3.3\,$\mu$m observation and KVN observation is different.}
\tablenotetext{\ddag}{For sources with CO non-detection, redshift values are from the literature that contain $L_{3.3}$ values.}
\end{deluxetable*}

\tabletypesize{\scriptsize}
\begin{deluxetable*}{lcccl}
    \tablecaption{ \label{tab:literature} Literature CO observations }
    \tablehead{ \colhead{Literature} & \colhead{CO line} & \colhead{N$_{3.3}$\tablenotemark{a}} & \colhead{Telescope} & \colhead{Targets description} }
    \startdata
    \citet{1995ApJS...98..219Y} & (1$-$0) & 22 & FCRAO (14\,m, SD) & Nearby galaxies \\     
    \citet{2015PASJ...67...36M} & (1$-$0) & 2 & Nobeyama (45\,m, SD) & Normal star-forming galaxies ($0.1<z<0.2$)  \\     
    \citet{2017ApJ...844...96Y} & (1$-$0) & 29 & Nobeyama (45\,m, SD) & LIRGs and ULIRGs \\     
    \citet{2019MNRAS.482.1618C} & (1$-$0) & 6 & IRAM (30\,m, SD) & Star-forming galaxies ($0.03<z<0.28$)   \\
    \citet{2019AandA...628A..71H} & (1$-$0) & 23 & IRAM (30\,m, SD) & LIRGs and ULIRGs  \\   
    \cite{2023AandA...673A..13M} & (1$-$0) & 10 & ALMA & Local (U)LIRGs  \\   
    \citet{2015AandA...578A..95I} & (2$-$1) & 13 & JCMT (15\,m, SD) & (U)LIRGs and starburst galaxies \\
    \citet{2020ApJS..247...15S} & (2$-$1) & 1 & ALMA & PG quasars  \\    
    \citet{2021ApJS..257...64K} & (2$-$1) &  3 & ALMA & Hard X-ray selected AGN   \\ 
    \cite{2022AandA...668A..45L} & (2$-$1) & 13 & ALMA & Nearby ULIRG  \\    
    \cite{2022AandA...664A..60B} & (2$-$1) & 1 & ALMA & Local ($z<0.02$) LIRGs  \\    
    \cite{2023AandA...673A..13M} & (2$-$1) &  2 & APEX (12\,m, SD) & Local (U)LIRGs \\   
    \citet{2015AandA...578A..95I} & (4$-$3), (7$-$6) & 16 & Herschel & (U)LIRGs and starburst galaxies \\
    \citet{2017ApJS..230....1L} & (4$-$3), (5$-$4), (6$-$5), (7$-$6) & 59 & Herschel & Local LIRGs  \\
    \enddata
\tablenotetext{a}{Number of galaxies where 3.3\,$\mu$m flux, not upper limit, is available}
\end{deluxetable*}

\tabletypesize{\normalsize}
\begin{deluxetable*}{lcccclc}
    \tablecaption{\label{tab:fit} Linear scaling relations between the $3.3\,\mu$m PAH, IR, and CO(1$-$0) luminosities}
    \tablehead{\colhead{$x$} & \colhead{$y$} & \colhead{$\alpha$} & \colhead{$\beta$} & \colhead{$\sigma^2$} & \colhead{Sample} & \colhead{N$_\mathrm{source}$}
    }
    \startdata
    $L_{3.3}$ [L$_\odot$] & $L_\mathrm{IR}$ [L$_\odot$] &  $0.80\pm0.04$ & $-0.85\pm0.51$ & 0.132 & all & 180 \\
    $L_{3.3}$ [L$_\odot$] & $L_\mathrm{IR}$ [L$_\odot$] &  $0.78\pm0.04$ & $-0.54\pm0.50$ & 0.102 & detection only & 150 \\
    $L_{3.3}$ [L$_\odot$] & $L_\mathrm{IR}$ [L$_\odot$] &  $0.75\pm0.05$ & $-0.24\pm0.54$ & 0.081 & non-AGN & 113 \\
    \hline
    \hline
    $L_\mathrm{CO(1-0)}^\prime$ [K km s$^{-1}$ pc$^2$] & $L_{3.3}$ [L$_\odot$] & $0.85\pm0.12$ & $0.42\pm1.13$ & 0.056 & KVN observations (all) & 49 \\
    $L_\mathrm{CO(1-0)}^\prime$ [K km s$^{-1}$ pc$^2$] & $L_{3.3}$ [L$_\odot$] & $0.97\pm0.14$ & $-0.66\pm1.30$ & 0.057 & KVN observations (detection only) & 44 \\
    \hline
    $L_\mathrm{CO(1-0)}^\prime$ [K km s$^{-1}$ pc$^2$] & $L_{3.3}$ [L$_\odot$] & $1.29\pm0.11$ & $-4.04\pm1.08$ & 0.068 & literature (all) & 86 \\
    $L_\mathrm{CO(1-0)}^\prime$ [K km s$^{-1}$ pc$^2$] & $L_{3.3}$ [L$_\odot$] & $1.09\pm0.11$ & $-2.11\pm1.00$ & 0.054 & literature (detection only) & 70 \\
    $L_\mathrm{CO(1-0)}^\prime$ [K km s$^{-1}$ pc$^2$] & $L_{3.3}$ [L$_\odot$] & $1.01\pm0.12$ & $-1.26\pm1.13$ & 0.033 & literature (subsample\tablenotemark{\dag}) & 49 \\
    \hline
    $L_\mathrm{CO(1-0)}^\prime$ [K km s$^{-1}$ pc$^2$] & $L_{3.3}$ [L$_\odot$] & $1.10\pm0.08$ & $-2.08\pm0.75$ & 0.063 & KVN+literature (all) & 135 \\
    $L_\mathrm{CO(1-0)}^\prime$ [K km s$^{-1}$ pc$^2$] & $L_{3.3}$ [L$_\odot$] & $1.00\pm0.07$ & $-1.10\pm0.70$ & 0.055 & KVN+literature (detection only) & 114 \\
    \hline
    \hline
    $L_\mathrm{CO(1-0)}^\prime$ [K km s$^{-1}$ pc$^2$] & $L_{3.3}$ [L$_\odot$] & $1.56\pm0.84$ & $-6.39\pm7.85$ & 0.131 & KVN observations (AGN) & 9 \\
    $L_\mathrm{CO(1-0)}^\prime$ [K km s$^{-1}$ pc$^2$] & $L_{3.3}$ [L$_\odot$] & $0.90\pm0.15$ & $0.07\pm1.34$ & 0.051 & KVN observations (non-AGN) & 35 \\
    \hline
    $L_\mathrm{CO(1-0)}^\prime$ [K km s$^{-1}$ pc$^2$] & $L_{3.3}$ [L$_\odot$] & $1.37\pm0.20$ & $-4.82\pm1.87$ & 0.069 & KVN+literature (AGN) & 25 \\
    $L_\mathrm{CO(1-0)}^\prime$ [K km s$^{-1}$ pc$^2$] & $L_{3.3}$ [L$_\odot$] & $0.87\pm0.08$ & $0.18\pm0.72$ & 0.048 & KVN+literature (non-AGN) & 89 \\ 
    \enddata
    \tablecomments{Linear scaling relations are found in the form of $\mathrm{log}\,y=\alpha\times\mathrm{log}\,x +\beta$.}
    \tablenotetext{\dag}{Sources from \citet{2020ApJ...905...55L} for which CO(1$-$0) SD observations are available.}
\end{deluxetable*}

\begin{acknowledgments}
The authors would like to thank the referee for providing helpful suggestions to improve the manuscript.
The KVN is a facility operated by the Korea Astronomy and Space Science Institute (KASI).
This work was supported by the National Research Foundation of Korea (NRF) grant 
funded by the Korea government (MSIT; Nos. 2022R1A4A3031306, 2022R1A6A3A01085930, RS-2024-00349364, 2021R1C1C1013580). 
\end{acknowledgments}

%

\vspace{5mm}
\facilities{KVN, AKARI (IRC and FIS), 
GALEX, SDSS, 2MASS, UKIDSS, WISE, Herschel (SPIRE), IRAS}


\software{Astropy \citep{2013A&A...558A..33A,2018AJ....156..123A},
\textsc{cigale} \citep{2009AandA...507.1793N,2019A&A...622A.103B},
GILDAS/CLASS \citep{2013ascl.soft05010G},
\texttt{linmix} \citep{2007ApJ...665.1489K},
matplotlib \citep{2007CSE.....9...90H},
NumPy \citep{2011CSE....13b..22V}.
}







\bibliography{ms_KVN}{}
\bibliographystyle{aasjournal}



\end{document}